\documentclass[twocolumn,twocolappendix]{aastex701}    %linenumbers
 
\hypersetup{linkcolor=red,citecolor=blue,urlcolor=magenta}

\usepackage{newtxtext,newtxmath} 
 
\newcommand{\hunit}{km s$^{-1}$ Mpc$^{-1}$} 

\newcommand{\kmps}{km s$^{-1}$}
\newcommand{\remaj}{\textit{$R_{\rm e}^{\rm maj}$}}
\newcommand{\as}{$^{\prime \prime}$}

\newcommand{\lmr}{\textit{$\lambda_{R}$}}
   
\newcommand{\newadd}[1]{\textcolor{red}{{#1}}}

\NewDocumentCommand\michele{O{Rewritten}m}{\textsf{\color{green}[Michele: #1]\(\rightarrow\)[``\emph{#2}'']}}

\begin{document}

\title{\Large MAGNUS I: A MUSE-DEEP Sample of Early-type Galaxies at Intermediate Redshift}

%use [] before {} to provide ORCID id info for authors
\author[0000-0002-8593-7243]{Pritom Mozumdar}%\thanks{corresponding author:pmozumdar@astro.ucla.edu} 
\email{pmozumdar@astro.ucla.edu}
\affiliation{Department of Physics and Astronomy, University of California, Los Angeles, CA 90095, USA}
\affiliation{Department of Physics and Astronomy, University of California, Davis, 1 Shields Ave., Davis, CA 95616, USA}

\author[0000-0002-1283-8420]{Michele Cappellari}
\email{michele.cappellari@physics.ox.ac.uk}
\affiliation{Sub-Department of Astrophysics, Department of Physics, University of Oxford, Denys Wilkinson Building, Keble Road, Oxford, OX1 3RH, UK}

% \collaboration{20}{(AAS Journals Data Editors)} 

\author[0000-0002-4030-5461]{Christopher D. Fassnacht}
\email{cdfassnacht@ucdavis.edu}
\affiliation{Department of Physics and Astronomy, University of California, Davis, 1 Shields Ave., Davis, CA 95616, USA}

\author[0000-0002-8460-0390]{Tommaso Treu}
\email{tommaso.treu@physics.ucla.edu}
\affiliation{Department of Physics and Astronomy, University of California, Los Angeles, CA 90095, USA}

\correspondingauthor{Pritom Mozumdar}
\email{pmozumdar@astro.ucla.edu}

\begin{abstract}
    We present a sample of 213 early-type galaxies (ETGs) at redshifts $0.25 < z < 0.75$. We combine deep integral-field spectroscopy from the MUSE-DEEP survey with high-resolution HST imaging to study the structure, kinematics, and stellar populations of these galaxies. We measure spatially resolved stellar kinematics and use the specific angular momentum proxy, $\lambda_R$, to classify galaxies into fast and slow rotators. The kinematic and photometric axes of fast rotators are generally well-aligned, similar to their local counterparts. We find that global stellar population properties, such as age, metallicity, and mass-to-light ratio ($M_*/L$), correlate strongly with the central velocity dispersion ($\sigma_\mathrm{e}$), following trends established for local ETGs. Slow rotators are typically more massive, have higher $\sigma_\mathrm{e}$, and are more metal-rich than fast rotators. We also find a slow rotator fraction consistent with local Universe samples. Our findings indicate that the fundamental structural, kinematic, and stellar population scaling relations of massive ETGs were already in place by $z \sim 0.75$, suggesting their evolutionary pathways have remained stable over the last $\sim 7$ Gyr.
\end{abstract}  


\section{Introduction} \label{sec:intro}
Understanding galaxy evolution over cosmic time is one of the key goals of observational astronomy. It is also crucial for understanding the formation of structure, dark matter-baryonic interactions, and large-scale cosmology. One can probe evolutionary trajectories by analyzing the mass distribution and kinematics and measuring the morphological, structural, and stellar population properties of galaxies (see the review of galaxy models by \citealp{Naab_Ostriker_2017}). In this regard, early-type galaxies (ETGs) are of particular interest as they represent the final stage of galaxy evolution, encapsulating the cumulative history of galaxy formation and transformation. ETGs also serve as valuable systems for studying the role of environmental processes in galaxy evolution. Mechanisms such as mergers, tidal interactions, and ram pressure stripping can profoundly alter their structural and kinematic properties, offering insight into how external influences shape their development (see the observational review by \citealp{Cappellari_2026_review}).

The advent of integral field unit (IFU) spectroscopy has revolutionized our understanding of ETGs by enabling spatially resolved studies of their kinematics, internal dynamics, and stellar populations across statistically significant samples \citep[see review of][]{Cappellari_review_2016}. Large IFU surveys, including SAURON \citep{SAURON_de_Zeeuw_2002}, ATLAS\textsuperscript{3D} \citep{ATLAS_I_Cappellari_2011}, CALIFA \citep{CALIFA_survey_2012}, SAMI \citep{SAMI_Bryant_2015}, and MaNGA \citep{MaNGA_survey_Bundy_2015}, have collectively established that ETGs in the local Universe fall into two kinematic classes, regular (or fast) and non-regular (or slow) rotators \citep{Kinematic_classification_Emsellem_2007, ATLAS_III_Emselem_2011, Vande_sande_SAMI_2017}, with a total density profile consistent with a nearly isothermal power law \citep{Cappellari_2015, Poci_2017, Li_2019, Dynpop_III_Zhu_2024}, a range of velocity anisotropies \citep{Cappellari_2007, Thomas_2009}, and stellar population properties that correlate with dynamical features, primarily with the central velocity dispersion $\sigma_{\mathrm{e}}$ \citep{ATLAS_XX_LOESS_Cappellari_2013b, McDermid_2015, Scott2017, Li2018_MaNGA, DynPop_II_Lu_2023}, consistent with the picture of archaeological downsizing in which more massive galaxies formed their stars earlier and more rapidly \citep{Cowie_1996, Thomas_2005, Gallazzi_2005}. \\

%provided comprehensive insights into the structural, kinematic, and stellar population properties of ETGs, shedding light on the key processes that drive their formation and evolution. For example, from these datasets, we have learned that ETGs mainly fall into two kinematic classes, regular (or fast rotators) and non-regular (or slow rotators) \citep{Kinematic_classification_Emsellem_2007, ATLAS_III_Emselem_2011}, and follow different evolutionary pathways, with a total density profile nearly isothermal \citep{Cappellari_2015, Poci_2017, Li_2019, Derkenne_2021, Dynpop_III_Zhu_2024}, contain a range of velocity anisotropies \citep{Cappellari_2007, Thomas_2009}, their stellar population properties correlate with dynamical features \citep{Dynpop_III_Zhu_2024, LEGAC_ppxf_Cappellari_2023}, and environmental mechanisms may have mild to significant influence on their mass and angular momentum content \citep{Munoz_Lopez_2024, Derkenne_2023, DEugenio_2023}.

At intermediate (z $\sim$ 0.5) and higher redshifts (z $>$ 1), spectroscopic surveys such as LEGA-C \citep{LEGAC_survey_Wel_2016}, KMOS$^{\mathrm{3D}}$ \citep{Wisnioski_2015}, VANDELS \citep{Carnall_2019}, MAGPI \citep{MAGPI_survey_design_foster_2021}, and the JWST SUSPENSE program \citep{Slob_2025} have extended our understanding of ETG properties beyond the local Universe. Building on the pioneering work of measuring absorption line 
strengths and velocity dispersions in cluster ETGs \citep[e.g.,][]{Bender_1998, Kelson_2000, Jorgensen_2005}, studies based on these surveys have revealed, among other things, that the kinematic dichotomy between fast and slow rotators \citep[e.g.][]{vander_Marel_2007, vandaer_Wel_2008, LEGAC_Bezanson_2018, van_Houdt_2021}, the stellar population scaling relations with $\sigma_{\mathrm{e}}$ \citep{Gallazzi_2014, Choi_2014, Chauke_2018, Beverage_2021, Beverage_2023, LEGAC_ppxf_Cappellari_2023, Gallazzi_2026a}, and the downsizing trends are broadly in place at $z \sim 1$. However, these studies have largely relied on slit-based spectroscopy or small IFU samples, and are worth revisiting with a comprehensive, homogeneous IFU-based sample of massive ETGs spanning a broad intermediate redshift range. \\

An IFU-based sample would also enable a range of analyses that are not fully achievable with slit-based data alone. Two-dimensional kinematic maps provide robust kinematic classification into fast and slow rotators, measurement of the kinematic--photometric misalignment angle \citep{pafit_Krajnovic_2006}, and the construction of dynamical models that constrain the mass distribution and orbital structure of galaxies \citep{Derkenne_2023, Dynpop_I_Zhu_2023}. Spatially resolved measurements of stellar population properties and star formation histories reveal how stellar populations vary within individual galaxies, offering vital constraints on their evolutionary pathways and the relative roles of quenching, mergers, and environmental processes in shaping them \citep{DynPop_II_Lu_2023, Dynpop_III_Zhu_2024, Mun_2024, Koller_2024}. Together, these capabilities are essential for addressing open questions at intermediate redshifts, including the evolution of rotational support \citep{LEGAC_Bezanson_2018}, the evolution of the total mass density slope \citep[e.g.,][]{Derkenne_2021, Sahu_AGEL_2024}, and the assembly history of massive ETGs across cosmic time. A homogeneous analysis of a large sample spanning a broad redshift range, with data quality and analysis methods consistent with local IFU surveys, is therefore necessary to advance our understanding of the evolutionary processes shaping massive ETGs over the last $\sim$7~Gyr. \\

\begin{figure*}
\centering
    \includegraphics[width=\textwidth]{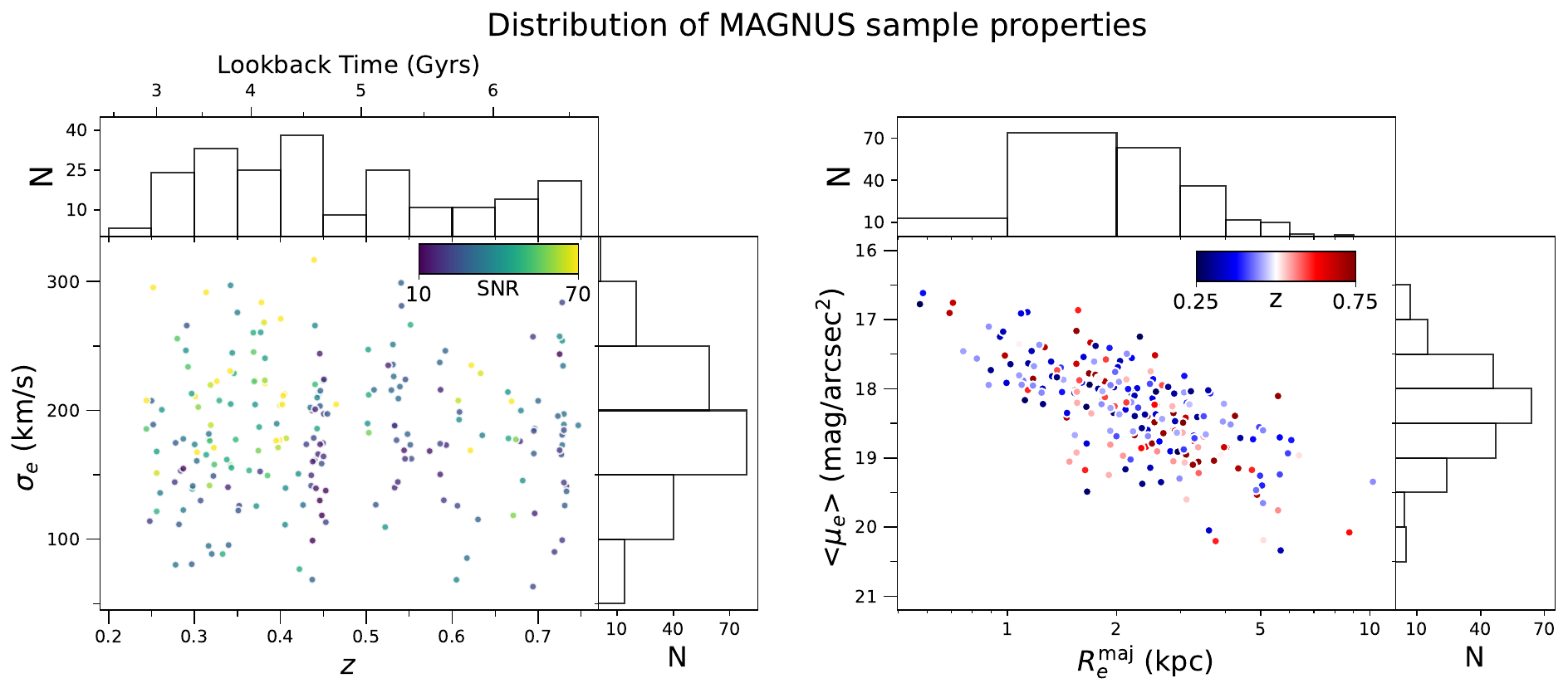}%width=\textwidth
    \caption{\label{fig:muse_sample_distribution} Distribution of MAGNUS sample properties. Left: integrated velocity dispersion, $\sigma_\mathrm{e}$, of the galaxies within the half-light isophote as a function of redshift, color-coded by the SNR per \AA\ of the central spectrum. Right: average surface brightness in the F814W band, $\langle\mu_e\rangle$, as a function of semi-major axis, \remaj, color-coded by redshift. Both quantities were measured within the elliptical half-light isophote and using the HST image of the galaxies.
    }
\end{figure*}

To accomplish these goals, one requires a sample of massive ETGs with high-quality IFU data favorable to extracting resolved stellar kinematics up to a few effective radii. The ESO archival MUSE-DEEP\footnote{\dataset[doi: 10.18727/ARCHIVE/42]{https://doi.org/10.18727/ARCHIVE/42}} collection \citep{ESO2017}, which combines data from single or multiple MUSE programs at specific sky locations into a single high-SNR IFU datacube, provides an ideal resource for this purpose. Utilizing this dataset, we compiled a sample of ETGs spanning the redshift range $0.25 < z < 0.75$ with spatially resolved stellar kinematics extending to several effective radii. \\

%data observed with the Multi-Unit Spectroscopic Explorer (MUSE) provides a significant source of IFU data with high signal-to-noise ratios (SNRs) for massive ETGs. The MUSE-DEEP\footnote{\url{https://doi.org/10.18727/archive/42}} collection \citep{ESO2017} combines observed data from single or multiple MUSE programs at specific sky locations into a single IFU datacube, achieving very high effective exposure. 
%\michele[]{Can you comment on when was the sample updated and how complete is this sample?}.
%Utilizing this high-SNR data, we compiled a sample of ETGs within the redshift range of 0.25-0.75 with resolved stellar kinematics extending to several effective radii.

In this paper, the first one of the 
%\michele[]{This name does not include ETGs, which is a key characteristic of the sample. Unless we remove statements that we specifically select ETGs. The name must include at least (i) MUSE, (ii) ETGs, (iii) Comsology/evolution/high-z. AI can give you many good suggestions.} 
`Muse-deep Analysis of Galaxy kinematics and dyNamical evolUtion across redShift' (\newadd{MAGNUS}) series, we will introduce the sample, describe the data extraction process, and present the global kinematic, morphological, and stellar population properties. In subsequent papers, we will use the MUSE data along with Hubble Space Telescope (HST) observations to perform evolutionary analysis, study possible environmental effects, and incorporate dynamical constraints from ETGs into time-delay cosmography \citep{TDCOSMO_V, TDCOSMO_milestone_2025}. This paper is organized as follows. \autoref{sec:sample} details the sample selection from the MUSE-DEEP datacubes. \autoref{sec:data} presents the extraction of stellar kinematics and photometry, including the methods for measuring integrated and resolved kinematics as well as surface brightness modeling. \autoref{sec:data_analysis} describes the measurement of structural, kinematic, and stellar population properties. \autoref{sec:results_discussion} presents the main results, including the kinematic classification and kinematic-photometric misalignment angle of the sample (\autoref{sec:result_kin_classifiation}--\ref{sec:kin_misalignment}) and the stellar population scaling relations (\autoref{sec:stellar_pop}), discussed throughout in comparison with local Universe and intermediate-redshift samples. The key findings are summarized in \autoref{sec:conclusion}. The kinematic maps and sample catalog for the MAGNUS sample are provided in the Appendix. Throughout this paper, we assume a concordance flat $\Lambda$CDM cosmology with $H_0 = 70.0$ \hunit\ and $\Omega_{\rm m} = 0.3$ when necessary.

\section{MAGNUS sample selection} \label{sec:sample}

MUSE is an integral-field spectrograph located at the Very Large Telescope (VLT) under the European Southern Observatory (ESO). This spectrograph provides a spatial sampling of 0.2\as over a 1$^\prime \times$ 1$^\prime$ field-of-view (FOV) in the wide-field mode assisted with an adaptive optics (AO) system \citep{MUSE_Bacon_2010}. It covers a wavelength range of 4800-9300 \r{A} with 1.25 \r{A} reciprocal dispersion and provides an average spectral resolution of R $\sim$ 2000 corresponding to an instrumental dispersion of $\sigma_{\text{inst}} \sim$ 65 km$\mathrm{s}^{-1}$ at around 5000 \AA. To conduct detailed analysis %kinematic, dynamical, and stellar population properties 
of the massive ETGs at intermediate redshifts, comparable to the ones done for local Universe ETGs, we need IFU data with high depth and resolution, and the MUSE-DEEP program is a natural choice in this regard. However, as no unified catalog is available for objects covered by the publicly available MUSE-DEEP datacubes, accessible in the ESO Science Archive Facility\footnote{\hyperlink{https://archive.eso.org/scienceportal/home}{https://archive.eso.org/scienceportal/home}} {\citep{ESO2017}}, we have to check each datacube manually. After initial inspection, we selected 35 datacubes containing at least 4-6 ETGs at the target redshift, and accompanied by publicly available HST imaging of the corresponding fields for detailed analysis.

The selected datacubes comprise five of the six frontier field galaxy clusters, Abell S1063, Abell 2744, Abell 370, MACS J0416, and MACS J1149 \citep{Lotz_2017}, part of the Hubble Ultra Deep Field (HUDF) \citep{Bacon_2017} and COSMOS field \citep{Scoville_2007}. Other well-known galaxy clusters and fields covered by these datacubes are Abell 1835, Abell 2895, Bullet cluster, MACS J0257, MACS J0940, MACS J1206, MACS J2129, MACS J2214, MS 0451-03, RX J1347–1145, RX J2129+0005, SDSS J1029+2623, and SMACS 2131. The complete list of galaxy clusters and fields, corresponding ESO archive ID of the datacubes, MUSE Program IDs of the observations used to compile the datacubes, effective exposure, number of galaxies selected from each cube, and reference of the catalogs are reported in \autoref{table:datacube_sum}. Several of the datacubes included in this study have previously been used in other published works to gather a sample of ETGs to explore various aspects of galaxy evolution \citep[e.g.][]{Munoz_Lopez_2024, Derkenne_2021}. However, due to difference in selection criteria, our ETG sample does not fully coincide with those from earlier works. The primary contribution of this work is the assembly of a large ETG sample drawn from a broad set of MUSE-DEEP datacubes, coupled with a consistent and systematic approach to kinematic extraction. Naturally, we only considered the portion of the above-mentioned field or clusters covered by MUSE-DEEP Cubes.

In most cases, the selected datacubes contain galaxy clusters or fields with published catalogs providing galaxy coordinates and redshifts. We used this information to identify galaxies within our desired redshift range. We note that the sample galaxies in the cluster fields are almost exclusively in the foreground, with a handful of systems at higher redshift by at most $0.05$. Therefore we do not need to worry about lensing magnification for this analysis. If a catalog is not available, then we measured galaxy redshifts using their spectra. From this redshift-restricted sample, we then selected 213 galaxies following two criteria. First, a galaxy should be sufficiently isolated to ensure that the measured kinematics and photometric features would have minimum contamination from surrounding galaxies. A galaxy was considered sufficiently isolated if no other galaxy existed within an annular region extending 1-2\as\ beyond its isophotal boundary. Second, the projected image of the galaxy in the sky plane must be spatially large enough ($>1$\as along the major axis) to extract resolved stellar kinematics up to a few effective radii. In general, we aimed for big, bright, and isolated galaxies for uncontaminated kinematics and photometric data. Although we implemented the above criteria as strictly as possible, there were some exceptions. \autoref{fig:muse_sample_distribution} shows the distribution of the final MAGNUS sample in terms of redshift, integrated velocity dispersion $\sigma_\mathrm{e}$, average surface brightness in the F814W band %\michele[]{Check my definition of $\mu_\mathrm{e}$.}
$\langle\mu_\mathrm{e}\rangle\equiv L/(2\pi R_\mathrm{e}^2)$, with $L$ the galaxy total luminosity and semi-major axis \remaj\ within the elliptical half-light isophote. The measurement process of these properties will be described in later sections. The sample contains a relatively higher number of galaxies in the redshift range 0.3--0.45, although other redshift bins also host comparable numbers of objects. The $\sigma_\mathrm{e}$ of the sample spans a range of approximately 100--300 \kmps\ with a peak near 180 \kmps. In terms of surface brightness, the brightest galaxies in the sample reach values around 16.5 mag/arcsec$^2$, while the faintest ones are down to 20.5 mag/arcsec$^2$ in the HST F814W band. The majority of galaxies ($\sim$ 75\%) fall within 17.5-19 mag/arcsec$^2$. The largest galaxies in the sample have a semi-major axis of around 10 kpc, while the smallest ones exhibit \remaj\ values as low as 0.6 kpc. Most galaxies ($\sim$ 80\%) have a \remaj\ between 1 and 4 kpc. The coordinates, redshifts, and other measured properties will be publicly available in electronic format after the acceptance of this manuscript.
%% Describe the sample in terms of the properties in Figure 1.
%From Figure \ref{fig:muse_sample_distribution} one can see that though the redshift bin from 0.3-0.45 has the maximum number of galaxies, other redshift bins still have a statistically comparable number of galaxies.  galaxies The associated color map in this plot provides redshifts of the galaxies.
% Provide more info about the sample. The median redshift of the sample is $z \sim$ 0.45, with 16 the and 84 the percentile is y and z. Also, same for velocity dispersion, size and surface brightness\\

%----------------------start of datacube info table------------
% \renewcommand{\arraystretch}{1.3}
\begin{deluxetable*}{c c c c c c}
    \tablewidth{\textwidth}
    \tablecaption{\label{table:datacube_sum}Summary of the MUSE-DEEP datacubes used in this work. The ESO archive ID of the datacubes, name of the galaxy cluster or field covered, MUSE program IDs of the observations used to compile the respective deep cube, effective exposure, number of galaxies collected from the cube, and reference of catalogs are listed in the table. The catalogs were used to identify galaxies in the MUSE white and HST images. } 
    \tablehead{\colhead{Cluster/Field} & \colhead{ESO Archieve ID} & \colhead{MUSE Program ID} & \colhead{Effective} & \colhead{Number of } &  Catalog\\   
    &  &  & \colhead{exposure} &  \colhead{galaxies} &  reference\\
               &  &  & \colhead{(sec)} &  \colhead{collected} & }
\startdata
A1063 & ADP.2017-03-28T12:46:01.331 & 095.A-0653(A) & 15885.0 & 7 & \citet{gal_cluster_catalog_A1063_MACS1149_Tortorelli_2018}\\
A1835 & ADP.2019-01-02T09:51:36.335 & 097.A-0909(A) & 11178.0 & 11 & \\ 
A2744 & ADP.2017-03-24T12:14:09.100 & 095.A-0181, 096.A-0496,  & 16345.0 & 12 & \citet{A2744_Mahler_2018} \\ 
         &                             & 094.A-0115  &         &    &  \\
A2895 & ADP.2019-10-18T09:34:16.487 & 299.A-5028, 60.A-9195,  & 16134.0 & 1 & \\
         &                             & 0102.B-0741  &         &    & \\
A370 & ADP.2017-06-06T13:13:38.674 & 096.A-0710, 094.A-0115 & 13655.0 & 18 & \citet{A370_catalog_Lagattuta_2019}\\ 
BULLET & ADP.2017-03-24T12:23:57.159 & 094.A-0115(B) & 6766.0 & 5 & \citet{galaxycluster_catalogue_Richard_2021}\\ 
CLJ1449 & ADP.2022-08-31T14:19:53.687 & 105.20C9.001 & 27780.0 & 2 & \\ 
COSMOSGR28 & ADP.2017-03-24T15:17:56.787 & 095.A-0118, 094.A-0247 & 22709.0 & 3 & \citet{COSMOS_field_catalog_Weaver_2022}\\
COSMOSGR32 & ADP.2018-04-13T08:06:56.375 & 097.A-0254, 0100.A-0607 & 13650.0 & 8 & \citet{COSMOS_field_catalog_Weaver_2022}\\ 
           & ADP.2019-01-02T10:06:51.914 & 0101.A-0282(A) & 10645.0 & 5 &  \\
COSMOSGR35 & ADP.2019-10-07T09:42:38.927 & 0103.A-0563, 0102.A-0327 & 14503.0 & 4 & \citet{COSMOS_field_catalog_Weaver_2022}\\ 
COSMOSGR83 & ADP.2017-03-24T15:17:56.840 & 094.A-0247, 096.A-0596 & 20413.0 & 2 & \citet{COSMOS_field_catalog_Weaver_2022}\\ 
COSMOSGR84 & ADP.2019-09-02T08:03:16.118 & 0102.A-0327(A) & 12302.0 & 6 & \citet{COSMOS_field_catalog_Weaver_2022}\\ 
COSMOSGR87 & ADP.2019-10-07T09:42:38.907 & 0103.A-0563(A) & 8563.0 & 2 & \citet{COSMOS_field_catalog_Weaver_2022} \\ 
COSMOSGR172 & ADP.2019-10-07T09:42:38.959 & 0103.A-0563(A) & 14553.0 & 2 & \citet{COSMOS_field_catalog_Weaver_2022}\\ 
HE0238 & ADP.2017-03-24T13:24:09.645 & 096.A-0222, 094.A-0131 & 24675.0 & 4 & \\ 
HUDF & ADP.2017-03-24T13:59:14.977 & 095.A-0010, 096.A-0045, & 31844.0 & 2 & \citet{HUDF_catalog_Inami_2017}\\ 
         &                             &  094.A-0289  &         &    & \\
     & ADP.2017-03-24T13:59:14.871 & 095.A-0010, 096.A-0045,  & 29993.0 & 1 & \citet{HUDF_catalog_Inami_2017}\\ 
         &                             & 094.A-0289  &         &    & \\
     & ADP.2019-11-20T14:57:13.894 & 1101.A-0127(A) & 321693.0 & 3 & \citet{HUDF_catalog_Inami_2017}\\ 
LAB1 & ADP.2017-03-28T12:26:19.431 & 095.A-0570(A) & 12417.0 & 1 & \\ 
MACS0257 & ADP.2019-11-25T09:40:00.922 & 0103.A-0157, 0100.A-0249 & 21304.0 & 7 &\citet{galaxycluster_catalogue_Richard_2021}\\ 
MACS0416 & ADP.2019-10-09T11:36:01.797 & 0100.A-0763(A) & 39545.0 & 11 & \citet{gal_catalog_MACS0416_Vanzella_2021}\\
         & ADP.2017-03-24T16:19:17.624 & 094.A-0525(A) & 36113.0 & 1 & \citet{gal_catalog_MACS0416_Vanzella_2021}\\
MACS0940 & ADP.2019-01-02T09:02:39.474 & 0101.A-0506, 0100.A-0249 & 21695.0 & 7 &\citet{galaxycluster_catalogue_Richard_2021}\\ 
MACS1149 & ADP.2017-03-24T16:26:09.634 & 294.A-5032(A) & 15282.0 & 12 & \citet{gal_cluster_catalog_A1063_MACS1149_Tortorelli_2018}\\ 
MACS1206 & ADP.2017-06-19T11:32:26.411 & 095.A-0181, 097.A-0269 & 17478.0 & 20 & \citet{galaxycluster_catalogue_Richard_2021}\\
MACS2129 & ADP.2017-03-27T15:26:07.717 & 095.A-0525(A) & 8802.0 & 8 & \citet{gal_catalog_MS0451_MACS2129_Jauzac_2021}\\ 
MACS2214 & ADP.2020-03-30T10:35:44.574 & 0101.A-0506, 099.A-0292, & 23976.0 & 4 &\citet{galaxycluster_catalogue_Richard_2021}\\ 
         &                             & 0103.A-0157,0104.A-0489  &         &    &\citet{galaxycluster_catalogue_Richard_2021}\\
MS0451 & ADP.2017-06-07T07:24:11.988 & 096.A-0105(A) & 8478.0 & 9 & \citet{gal_catalog_MS0451_MACS2129_Jauzac_2021}\\ 
NGC1052 & ADP.2018-12-20T09:53:26.537 & 2101.B-5008, 2101.B-5053 & 16138.0 & 1 & \\ 
RXJ1347 & ADP.2019-01-02T09:51:36.307 & 097.A-0909(A) & 8526.0 & 13 & \citet{galaxycluster_catalogue_Richard_2021}\\ 
RXJ2129 & ADP.2017-12-14T12:30:03.217 & 097.A-0909(A) & 11444.0 & 1 & \\ 
RXS0437 & ADP.2021-04-29T08:02:29.347 & 106.21AD.001 & 7711.0 & 11 & \\
SDSSJ1029 & ADP.2019-08-30T07:41:35.874 & 0102.A-0642(A) & 13387.0 & 3 & \\ 
SMACS2131 & ADP.2019-11-25T09:40:00.887 & 0101.A-0506, 0103.A-0157,  & 22784.0 & 6 & \citet{galaxycluster_catalogue_Richard_2021}\\ 
         &                             & 0102.A-0135  &         &    & \\
\enddata
\end{deluxetable*}

%------------------end of datacube info table-------------------

\section{Kinematics extraction and photometry} \label{sec:data}
The following section describes how 2D stellar kinematics maps for individual galaxies were extracted from the MUSE-DEEP datacubes. It also includes a brief overview of how HST images were used to model the surface brightness of each galaxy. 

\subsection{Stellar kinematics}
\label{sec:stellar_kinematics}
First, we summarize the kinematics method and then describe how it was applied to measure integrated and resolved kinematics for all galaxies.

\subsubsection{Methodology of kinematics measurements}
\label{sec:kin_method}
We applied the recipe proposed by \citet[][hereafter KM25]{Knabel_Mozumdar_2025} to measure the integrated and resolved stellar kinematics, while estimating and minimizing the associated statistical and systematic uncertainties, as described in more details in \citet{Mozumdar_Knabel_2025}. KM25 showed that the usage of defective templates, e.g. those with flux calibration or telluric correction inaccuracies, can introduce significant bias in the kinematics measurements. Both KM25 and \citet{Mozumdar_Knabel_2025} found that the template mismatch is the main source of systematic uncertainty in stellar kinematics of early-type galaxies. Therefore, we have created three clean stellar libraries: Indo-US \citep{Indo_US_lib}, MILES \citep{MILES_library_2011}, and X-shooter Spectral Library (XSL) \citep{Verro_2022} using the list of clean stars provided in KM25. Some of the stars in these lists also contain defects in several wavelength regions, and KM25 further flagged these stars with the information of the problematic wavelength range. We checked for the presence of these defective regions in the clean templates within the wavelength range of 3600-4800 \AA. If defects are present, we also exclude these spectra from the clean library. The final clean Indo-US, MILES, and XSL libraries contain 989, 789, and 462 templates, respectively. We used these three clean libraries for all the kinematics measurements. The approximately constant full width at half maximum (FWHM) kinematic resolution of Indo-US, and MILES templates are 1.36 \AA\ and 2.5 \AA, respectively \citep{MILES_library_2011}, which correspond to $\sigma_{\text{t}} \approx$ 43 km s$^{-1}$ and 80 km s$^{-1}$, respectively, at 4000 \AA. The kinematic resolution of the XSL templates is $\sigma_{\rm t} \approx 13$ km s$^{-1}$ at the UVB region (3000--5560 \AA) \citep{Verro_2022}. We refer the reader to the original papers for a full description of the libraries. \\

To access the quality of a galaxy spectrum, we used the quantity mean SNR per \AA. This is estimated by measuring the mean SNR per pixel within the rest frame wavelength interval of 3990--4190 \AA\ and dividing by the square root of the number of \AA\ per pixel. Throughout this work, we have adopted a single rest-frame wavelength range of 3780-4600 \r{A} for extracting kinematics to ensure that galaxies from different redshifts contain the same absorption features such as $\mathrm{CN}$, $\mathrm{C}$a II$\mathrm{K}$ and  $\mathrm{H}$, and $\mathrm{G}$ band, etc. The cutoff of the adopted wavelength range on the blue side is imposed by both the instrument setup and the redshift, and on the red side by the redshift and to avoid contamination from strong skylines beyond 8000 \AA. \citet{Mozumdar_Knabel_2025} showed that for data and templates of sufficient quality, on average, the stellar velocity dispersion changes at the level of 1\%, due to dependence on absorption features and small changes (around 100-200 \AA) of the wavelength range. Thus, we further ignored systematic uncertainties from wavelength dependence. \\

Issues with galaxy spectra, such as improper flux calibration, variance in continuum strength, etc., may bias the kinematics fit significantly if additive and multiplicative polynomials are not introduced. Fits with lower polynomial orders than the optimal level are generally unstable, leave significant residuals, and result in disagreements between the template libraries. Around the optimal polynomial order, which depends on the wavelength range and quality of the flux calibration, the fits become stable and yield good agreement between the templates. If the polynomial order is too high, the polynomials start modifying the shape of the absorption lines, and the fit becomes unstable again. Using a subsample of MAGNUS galaxies, KM25 ran a grid of polynomials and found that additive and multiplicative polynomials of order 4 and 2, respectively, provide the least systematic scatter between templates for the sample. We therefore use the above polynomial orders for this dataset and ignore systematic uncertainties from the change of polynomial orders. \\

To measure the kinematics, we have used a penalized pixel fitting method implemented through the \textsc{pPXF} package \citep{Cappellari_2004, Cappellari_2017_ppxf, LEGAC_ppxf_Cappellari_2023}. The instrumental resolution or line spread function (LSF) of galaxies was set to 2.5 \r{A} (FWHM) \citep{MUSE_HUDF_kinematics_Adrien_2017}. Although the LSF of MUSE depends on the wavelength, it does not vary significantly within our adopted wavelength range to affect the kinematic measurements. If the templates have higher resolution than the redshift corrected instrumental resolution of the galaxy, a Gaussian kernel was used to broaden the templates and its FWHM was set to \(\mathrm{FWHM_{diff}^2=FWHM_{gal}^2 - FWHM_{temp}^2}\), the quadratic difference between the redshift-corrected FWHM resolution of MUSE for the respective galaxy and the template. If lower, then the measured velocity dispersion was post corrected to account for that. We excluded some wavelength regions from the fits at two stages. First, we masked the wavelength ranges containing gaps, if present. Next, using an iterative sigma clipping method, we detected wavelength ranges containing emission lines, instrument or extraction-related artifacts, cosmic rays, etc., if any, and also masked them from the fit.

We used all three clean libraries or a subset of the templates in the libraries to measure the velocity and velocity dispersion from a galaxy spectrum. The mean velocity dispersion $\overline {\sigma_{\rm{v}}}$, the associated systematic $\Delta \overline {\sigma_{\rm{v}}}$ and statistical uncertainty $\delta \overline{\sigma_{\rm{v}}}$ are then defined as weighted average such that

\begin{equation}
\overline{\sigma_{\rm{v}}} = \frac{\sum_{k} w_k \sigma_{\rm{v},k}}{\sum_k w_k}
\label{eq:mean}
\end{equation}
\begin{equation}
(\Delta \overline{\sigma_{\rm{v}}})^2 = \frac{\sum_{k} w_k (\sigma_{\rm{v},k}-\overline{\sigma_{\rm{v}}})^2}{\sum_k w_k},
\label{eq:sys_error}
\end{equation}
and
\begin{equation}
\delta \overline{\sigma_{\rm{v}}} = \sqrt{\frac{\sum_{k} w_k (\delta \sigma_{\rm{v},k})^2}{\sum_k w_k}},
\label{eq:stat_error}
\end{equation}

where the index $k$ runs through the template library, $\sigma_{\rm{v},k}$ is the velocity dispersion and $\delta \sigma_{\rm{v},k}$ is the associated statistical uncertainty estimated by \textsc{pPXF}, and $w_k$ is the weight assigned to a library. Similarly, by replacing $\sigma_{\rm{v},k}$ with $\rm{v}_k$ and $\delta \sigma_{\rm{v},k}$ with $\delta \rm{v}_k $, the above set of equations were also used to measure the weighted mean velocity $\overline {\rm{v}}$, and corresponding systematic $\Delta \overline {\rm{v}}$ and statistical uncertainty $\delta \overline{\rm{v}}$. For simplicity, we assigned equal weight to each library instead of the Bayesian Information criterion (BIC) determined weights recommended in KM25. 

%This method measures stellar kinematics by fitting a model to the observed galaxy spectrum in pixel space. The models are created using a weighted linear combination of the broadened stellar templates to which a sum of orthogonal polynomials is added. The additive polynomials are used to adjust the continuum shape of the templates during the fit.The wavelengths and instrumental FWHM resolutions of the galaxies have been shifted to the rest frame as prescribed in \citet{Cappellari_2017_ppxf}. 

\begin{figure*}
    \centering
    {\includegraphics[scale=0.53]{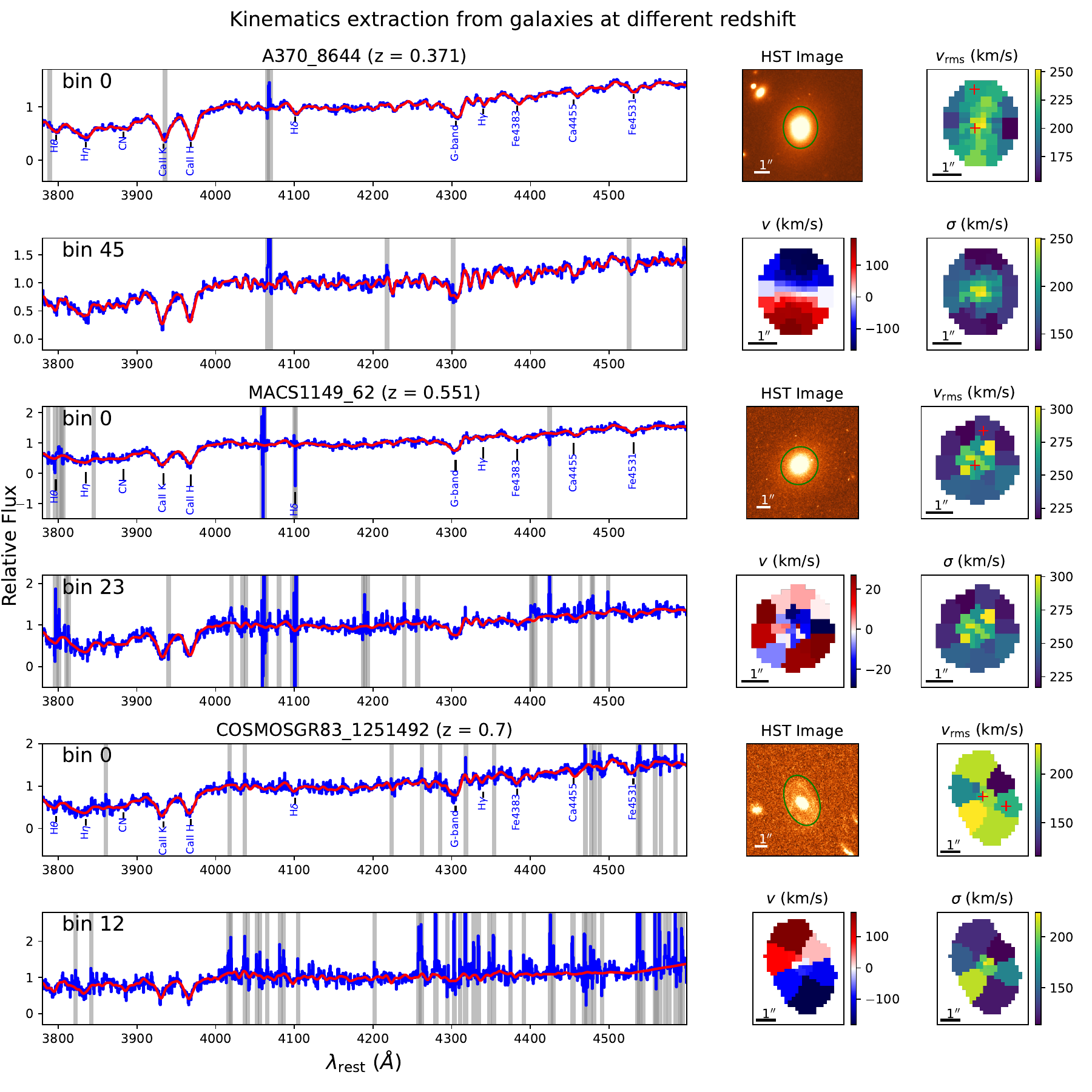}}%width=0.95\textwidth, height=0.7\textheight
    \caption{Example of kinematics extraction from galaxies at different redshifts. The galaxies are A370\_8644 ($z=0.371$, top), MACS1149\_62 ($z=0.551$, middle) and COSMOSGR83\_1251492 ($z=0.7$, bottom). In each case, two extracted spectra (blue) from two different Voronoi bins, one close to the galaxy center and the other close to the edge, are shown with kinematic fitting from \textsc{pPXF} (red). 
    %The green markers are residuals between data and the fitted model at each wavelength and 
    The gray regions mark the excluded wavelength range from the fitting. Additionally, the corresponding HST image of the galaxy, extracted velocity (median subtracted), velocity dispersion, and RMS velocity maps are shown. The green contour in each HST image shows the aperture of the extracted kinematic data relative to the observed surface brightness. The red markers in the vrms map shows the luminosity weighted coordinates the of the bins.}
    \label{fig:kin_measurement}
\end{figure*}

\subsubsection{Kinematics measurement}
The MUSE-DEEP collection provides reduced 3D data and associated variance cubes. The original MUSE data covering the same field from different programs are reduced using the MUSE data reduction pipeline \citep{MUSE_pipeline_Weilbacher_2020}. This pipeline\footnote{For more details, see \url{https://www.eso.org/rm/api/v1/public/releaseDescriptions/102}} removes the instrument signature, performs sky correction, and then resamples and combines the relevant data into a single cube. From the original MUSE-DEEP cubes, we collect a smaller individual cube for each galaxy using the MUSE Python Data Analysis Framework (MPDAF) \citep{MPDAF_Bacon_2016}. The following steps were then executed on the individual data cubes to measure the integrated and resolved kinematics. \\

A central spectrum was created by co-adding the spectra from all spaxels within one effective radius ($R_\mathrm{e}$) (see section \ref{sec:measure_structure}).
%\michele[It's somewhat unusual to use a different $R_e$ than the one defined in \autoref{sec:measure_structure}. Is there a specific reason for this choice? State in a sentence what this code does: e.g.\ fit Sersic?]{The $R_\mathrm{e}$ was measured from the white-light image of the data cube} utilizing the 'flux\_radius' function available at python implementation for Source Extraction and Photometry (SEP) \citep{SEP_Barbary_2016} package.
The central spectrum was then used to measure the redshift and the integrated line-of-sight (LOS) stellar velocity and velocity dispersion. All the templates in the respective clean libraries were provided during this fit. From this fit, we create a global
template set which consists of only those templates with nonzero weights assigned by \textsc{pPXF}. Later, we used this global template set to measure the resolved kinematics in each Voronoi bin of the corresponding galaxy.
%\michele[I tried to clarify, but I am not sure: did you use a single global template or all nonzero templates?]{The set of templates with nonzero weights from the initial pPXF fit to the central spectrum--referred to as the global template set--was subsequently used to measure the resolved kinematics in each Voronoi bin of the corresponding galaxy.}
%The set of templates with nonzero weights from the initial \textsc{pPXF} fit to the central spectrum--referred to as the global template set--was subsequently used to measure the resolved kinematics in each Voronoi bin of the corresponding galaxy.
The global template sets reduce bin-to-bin scatter in measured resolved kinematics, otherwise present if the full libraries were used. We choose to use the global template set instead of a single global template, created from the global template set but with fixed weights obtained during the fitting of the central spectrum \citep{Derkenne_2021, Shajib_2023}, to allow flexibility for a possible population gradient in the galaxy going from the central region to the outskirts. One can also create a different global template for different radial annuli by fitting the integrated spectra to the corresponding annuli to account for the population gradient \citep{Knabel_2024}. However, this method is not practical for the MAGNUS sample, as not all galaxies have sufficient spatial extent to implement this. This central spectrum was also used to study global stellar population properties such as age, chemical composition, mass-to-light ratio ($M_{\ast}/L$), etc. The median SNR of the central spectra of the MAGNUS sample is 35 per \AA. %\includegraphics{figures/mass_prop.pdf}\michele[]{Here and in various places: S/N per \AA} per \AA. \michele[]{Again: also give the more meaningful S/N per `velscale` interval, e.g. S/N per 60 \kmps\ interval.} \\

To measure resolved kinematics from IFU data, it is standard practice in the literature to first select spaxels of the galaxy above a certain SNR threshold, Voronoi bin the selected spaxels at a target SNR, and then measure the kinematics from the binned spectra. This process ensures that spaxels dominated by background noise are excluded, and the selected spaxels are spatially grouped in a way to enhance the SNR of the binned spectra for reliable kinematics fits. Due to the difference in effective exposure, redshift, and environment, the SNR level of the background in the datacubes varies significantly. Therefore, we first measured the background SNR level in the datacube and selected the spaxels with SNR 3$\sigma$ above the background level. We determined the background level by fitting a Gaussian to the histogram of the SNR of the spaxels in the datacube. The mean of the fitted Gaussian is considered the background level. The SNR of each spaxel is the mean SNR within the restframe wavelength range of 3990-4190 \r{A}. The median of the background level for the MAGNUS sample is 0.5 per pixel with a range of 0.1-3.4 per pixel, and the median of the 3$\sigma$ SNR threshold is 1 per pixel with a range of 0.5 - 5.5 per pixel. \\

We performed Voronoi binning on the selected spaxles using the \textsc{VorBin} Python package \citep{Voronoi_Cappellari2003}. As the data quality varies significantly across galaxies, a fixed target SNR would not have been appropriate for all cases. Setting the target SNR too high in low-SNR data would result in excessively large bins, thus erasing spatial structure or failing to reach the target SNR in large portions of the field. On the other hand, setting the threshold too low in high-SNR data would produce a very large number of bins, often consisting of single spaxels, leading to oversampled maps. To balance spatial resolution and spectral reliability, we chose the target SNR empirically for each galaxy within the range of 5-25 per \AA, 
%\michele[]{per \AA\ equivalent to S/N of xx per `velscale`},
ensuring that key absorption features remained measurable in the binned spectra while avoiding unnecessary bin fragmentation or spatial oversmoothing. In general, we chose a target SNR $\geq$ 5 for low-SNR galaxies such that the Voronoi map of the galaxy contains at least 10 bins, while a higher target SNR was used for high-SNR galaxies to reduce the number of bins. 

%Distribution of the MAGNUS sample in terms of the number of Voronoi bins and spatial extension (in units of \textit{$R_{\rm e}^{\rm maj}$}) is also shown in the right subplot of Fig. 1. The associated color map in this plot provides redshifts of the galaxies. The resolved kinematics of the galaxies were measured up to these spatial extensions. From the associated histograms in this plot, it is apparent that around two-thirds of the total sample has more than 20 bins and a kinematic extension within 1\textit{$R_{\rm e}^{\rm maj}$} to 3\textit{$R_{\rm e}^{\rm maj}$}.\\

The derived Voronoi map is used to coadd the spectra in the spaxels to obtain a single spectrum per bin. We then used the above-mentioned method and the global template sets to measure the stellar velocity and velocity dispersion from each binned spectrum, resulting in a velocity and velocity dispersion map for each galaxy. Examples of \textsc{pPXF} fitting to spectra from different bins of the same galaxy and the resulting kinematics maps are shown in \autoref{fig:kin_measurement} for three galaxies at different redshifts.
%\michele{This makes the measured kinematics more robust in the lower-\(S/N\) bins} of a galaxy.

% At this point, those galaxies with less than 10 bins were excluded from the sample. We have implemented the criteria of target SNR and minimum bin number to enhance the reliability of the constraints on the model parameters derived using resolved kinematics data. 

%----------------------start of HST image and catalog info table------------
% \renewcommand{\arraystretch}{1.3}
\begin{deluxetable*}{c c c c c c c}
    % \tablewidth{\textwidth}
    \tablecaption{\label{table:hst_sum}Information about archival HST images used in this work are provided in the following table. These HST data were obtained from the Mikulski Archive for Space Telescopes (MAST) at the Space Telescope Science Institute. The specific observations analyzed can be accessed via \dataset[doi: 10.17909/6de9-dg36]{https://doi.org/10.17909/6de9-dg36}.}
    %\footnote{\href{https://archive.stsci.edu/}{{https://archive.stsci.edu/}}}. } 
    \tablehead{\colhead{Cluster/Field} & Observation date & Total Exposure & Instrument & Filter & Proposal ID & PI \\   
    &  &  \colhead{(sec)} &  &  &  & }
\startdata
A1063 & 2015-10-18 & 5046.0 & ACS/WFC & F814W & 14037 & Jennifer Lotz\\ 
A1835 & 2010-06-26 & 4840.0 & ACS/WFC & F814W & 11591 & Jean-Paul Kneib \\ 
A1835 & 2019-05-08 & 2430.0 & ACS/WFC & F555W & 15661 & Michael McDonald\\ 
A2744 & 2014-07-01 & 5044.0 & ACS/WFC & F814W & 13495 & Jennifer Lotz\\
A2895 & 2006-08-23 & 1200.0 & ACS/WFC & F606W & 10881 & Graham Smith \\ 
A370 & 2016-01-20 & 5146.0 & ACS/WFC & F814W & 14038 &  Jennifer Lotz\\ 
Bullet & 2011-02-18 & 4480.0 & ACS/WFC & F814W & 11591 & Jean-Paul Kneib \\ 
CLJ1449 & 2013-05-20 & 700.0 & WFC3/UVIS & F606W & 12991 & Veronica Strazzullo\\ 
COSMOSGR28 & 2004-04-18 & 2028.0 & ACS/WFC & F814W & 9822 & Nicholas Scoville \\ 
COSMOSGR32 & 2004-05-03 & 2028.0 & ACS/WFC & F814W &  9822 & Nicholas Scoville \\ 
COSMOSGR35 & 2004-05-09 & 2028.0 & ACS/WFC & F814W & 9822 & Nicholas Scoville\\ 
COSMOSGR83 & 2003-11-25 & 2028.0 & ACS/WFC & F814W & 9822 & Nicholas Scoville\\ 
COSMOSGR84 & 2004-05-08 & 2028.0 & ACS/WFC & F814W & 9822 & Nicholas Scoville\\ 
COSMOSGR87 & 2011-06-05 & 480.0 &  ACS/WFC & F814W & 12328 & Van Dokkum \\ 
COSMOSGR172 & 2004-04-16 & 2028.0 & ACS/WFC & F814W & 9822 & Nicholas Scoville\\ 
HE0238 & 2017-06-19 & 2182.0 & ACS/WFC & F814W & 14660 & Lorrie Straka \\ 
HUDF & 2010-02-09 & 5332.0 & ACS/WFC & F814W & 11563 & Garth Illingworth \\ 
MACS0257 & 2011-06-16 & 1440.0 & ACS/WFC & F814W & 12166 &  Harald Ebeling \\ 
MACS0416 & 2014-01-07 & 5246.0 & ACS/WFC & F814W & 13496 & Jennifer Lotz\\ 
MACS0940 & 2019-05-18 & 7526.0 & ACS/WFC & F814W & 15696 & David Carton\\ 
MACS1149 & 2014-12-22 & 5308.0 & WFC3/UVIS & F814W & 13790 & Steve Rodney\\ 
MACS1206 & 2011-05-25 & 1091.0 & ACS/WFC & F814W & 12069 & Marc Postman\\ 
MACS2129 & 2003-09-11 & 4530.0 & ACS/WFC & F814W & 9722 & Harald Ebeling\\ 
MACS2214 & 2003-10-29 & 4560.0 & ACS/WFC & F814W &  9722 & Harald Ebeling\\ 
MS0451 & 2011-02-07 & 4880.0 & ACS/WFC & F814W & 11591 & Jean-Paul Kneib\\ 
NGC1052 & 2020-09-15 & 2080.0 & ACS/WFC & F814W & 15851 & Pieter van Dokkum\\ 
RXJ1347 & 2006-03-10 & 5280.0 & ACS/WFC & F814W & 10492 & Thomas Erben\\ 
RXS0437 & 2021-11-21 & 1200.0 & ACS/WFC & F814W & 16670 & Harald Ebeling\\ 
SDSSJ1029 & 2011-05-15 & 5276.0 & ACS/WFC & F814W & 12195 & Masamune Oguri \\ 
%SGASJ2111 & 2009-04-14 & 1100.0 & WFPC2 & F814W & 11974 & Sahar Allam\\ 
SMACS2131 & 2011-05-03 & 1440.0 & ACS/WFC & F814W & 12166 & Harald Ebeling\\ 
\enddata
\end{deluxetable*}
% \tablecomment{} 
%------------------end of HST image and catalog info table-------------------

\subsection{Photometry}
\label{sec:photometry}
We cut out individual images for galaxies, typically with 10--15\as$^2$ fields, from the HST images covering the same sky location as the MUSE-DEEP cubes. We measured the background level in an image by fitting a Gaussian to the histogram of the photon count in each pixel. The mean of the fitted Gaussian was adopted as the background level and subtracted from the image. We measured the surface brightness of each galaxy using its sky-subtracted image. For this purpose, we integrated the flux within its elliptical half-light isophote, corrected for cosmological surface brightness dimming, and converted to AB magnitudes using the photometric zeropoint of the F814W filter.

We also used sky-subtracted images to model the surface brightness of the galaxies as a series of 2D Gaussian profiles implemented by the python package \textsc{MgeFit} \citep{MGE_Cappellari_2002}. The MGE models were used to measure structural properties such as ellipticities, \textit{$\epsilon$}, semimajor axes, \remaj, and effective radii, \textit{$R_\mathrm{e}$}, etc., within the elliptical half-light isophotes of the galaxies. \autoref{table:hst_sum} summarizes the relevant information, such as exposure, instrument setup, and filter used, etc., about the publicly available archival HST images used for this sample. We have used HST images taken with an F814W filter for almost all galaxies. For those without an HST image with F814W filter, we used images taken with a F555W or F606W filter. 

%\michele[We can probably remove this from here as you already have all details about MGE later]{Details of the MGE modeling procedure are described in Mozumdar et al. (2024, in preparation) where the MGE models have been used as a light tracer in the Jeans's axisymmetric dynamical modeling of the resolved kinematics.}  

\section{Measurement of galaxy properties}
\label{sec:data_analysis}
This section discusses the measurement methods of different structural, kinematic, and morphological galaxy properties using the extracted stellar kinematics and photometry data. First, we describe how ellipticity, effective radii, a proxy for angular momentum, \lmr, etc., were measured. After that, we explain the stellar population synthesis methods used in measuring the age, metallicity, mass-to-light ratio, and stellar mass of each galaxy.

\subsection{Measuring size and ellipticity}\label{sec:measure_structure}
We have used the MGE models of the galaxy surface brightness to measure ellipticity, defined as \textit{$\epsilon$} $=1-b/a$, semi-major axis, \remaj\ and circularized effective radius, \textit{$R_{\mathrm{e}}$}. These quantities were measured within the elliptical isophote covering half of the total observed light. Here, the circularized effective radius satisfies the condition $\pi R_{\mathrm{e}}^2=A$, where $A$ is the area of the elliptical half-light isophote. The `mge\_half\_light\_isophote' function from the \textsc{JamPy} package \citep{Jampy_Cappellari_2008} has been used for this purpose. This function creates a synthetic galaxy image using the MGE model (only for one quadrant due to symmetry) and measures the light enclosed within a grid of isophotes along the MGE major axis. Then it detects the isophote containing half of the analytic MGE total light using linear interpolation. The structural properties are measured within this isophote \citep[for details see][sec.~3.3.1]{ATLAS_XV_Michele_2013}. 
%As a sanity check, we have also estimated the circular $R_{\mathrm{e}}$ of each galaxy from the respective sky-subtracted HST image using a growth curve method and compared it with the corresponding $R_{\mathrm{e}}$ from the MGE models. Though the physical definitions of these two are different, we expect them to be close to each other if the MGE models were true physical representations of the galaxy surface brightness. In \autoref{fig:Re_measurement} these two measurements are plotted in log-log space which confirms our prediction.
Throughout the paper, we have used \remaj\ as the physical size estimate of a galaxy, given its robustness to inclination effects \citep[fig.~5]{ATLAS_XV_Michele_2013}.
%as it is more physically robust and less dependent on inclination. 

\subsection{Measuring photometric position angle}
We measured the photometric position angle (PA) of a galaxy from its sky-subtracted HST image using the `find\_galaxy' routine in the \textsc{MgeFit} package. This function computes the galaxy's center and PA by diagonalizing the intensity-weighted moment of inertia of the light distribution over a region of connected pixels. For some galaxies, the measured PA may vary based on the number of pixels used, as the galaxy shape changes from rounder to flatter as it extends outward. In our analysis, we used all the connected pixels of the galaxy image that are 5$\sigma$ above the background using the `level' keyword. The $\sigma$ is the scale of the Gaussian profile fitted to the histogram of the photon counts in the pixels to estimate the background. 

%The uncertainty in the measured PA is estimated by changing the `level' from 4$\sigma$ to 6$\sigma$ in increments of 0.25$\sigma$, and computing the standard deviation of the resulting set of PA measurements. The median uncertainty in PA for the sample is x, with a median standard deviation of y-z.
%The aperture size is specified using the 'fraction' keyword, which sets what fraction of the image's connected pixels should be included. However, the measured PA may vary depending on the aperture size, as the galaxy shape may change from rounder to flatter as it extends outward. Since we aim to compare the photometric PA with the corresponding kinematic PA derived from the velocity map (more on this in sections \ref{sec:kin_PA} and \ref{sec:kin_misalignment}), for consistency, we restrict the photometric PA measurement to the same spatial extent as the kinematic map. We determined the aperture size of a kinematic map in the following way using the OpenCV library\footnote{\url{https://pypi.org/project/opencv-python/}} via its python interface (cv2). We created a binary image of the individual MUSE cube in which the spaxels used in Voronoi binning were separated from the others. This binary image was used to find the smallest contour enclosing all the Voronoi binned pixels using the `findContours' function, and then an ellipse was fit to this contour using the `fitEllipse' function. We used the fitted ellipse as the aperture to measure the photometric PA.
%Usually, Photometric PAs are often measured using larger apertures because photometric images typically cover a wider field than the kinematic data.

\subsection{Measuring kinematic properties}
\label{sec:kin_property}
The resolved stellar kinematics information enables us to measure important galaxy properties such as the angular momentum and kinematic PA of the major axis. These properties are essential to apply a kinematic classification of the ETGs. In the following sections, we describe how the kinematic properties are measured.
%It has been observed in the local universe that the slow rotators are increasingly rare in field environments and relatively abundant at the central location of the clusters or groups
\subsubsection{Proxy for specific angular momentum}
\label{sec:lamr}
As an approximation to the specific angular momentum, we used a luminosity-weighted quantity $\lambda_R$ \citep{Kinematic_classification_Emsellem_2007} defined as
\begin{equation}
\label{eq:lmr}
\lambda_R=\frac{\langle R|V|\rangle}{\left\langle R \sqrt{V^2+\sigma^2}\right\rangle}=\frac{\sum_{i=0}^N F_i R_i\left|V_i\right|}{\sum_{i=0}^N F_i R_i \sqrt{V_i^2+\sigma_i^2}},
\end{equation}
where $V_i$ is the stellar velocity, $\sigma_i$ is the velocity dispersion, $R_i$ is the distance from the galactic center, and $F_i$ is the flux of a spaxel within the adopted aperture. As in $V_i$ and $\sigma_i$, we have used the velocity and velocity dispersion of the Voronoi bin to which the spaxel belongs. We summed over all the spaxels within an elliptical aperture where the semi-major axis is set to twice that of \remaj. However, the aperture’s ellipticity and position angle were set based on measurements at the half-light isophote. \\

We adopted a different aperture from the usual circular aperture of 1$R_\mathrm{e}$ for two reasons. First, in any case, an elliptical aperture captures the change in velocity map along the major axis relatively better than a circular aperture, and this becomes more important for flatter galaxies. Second, we found that an elliptical aperture with a semi-major axis of 1\remaj\ contains less than 10\% of the total spaxels within the 2D kinematics map, for 71\% of our sample galaxies. However, for the adopted larger aperture, this happens only for 8\% galaxies. An in-depth comparison between $\lambda_R$ measurements within 1 and 2$R_{\rm e}$, from high-quality CALIFA IFS data, was presented in \citet{FalconBarroso2019}. They found that for fast rotator galaxies, $\lambda_{\rm 2Re}\approx1.19\lambda_{\rm Re}$. We also observed a similar effect that on average $\lambda_R$ increases by 30\% when we used a 2\remaj\ aperture compared to a 1\remaj\ aperture. In general, we found that the aperture size does not affect the quantitative classification of a galaxy. \\
%Check appendix \ref{apndx:aperture_lmr} for more discussion on the effect of aperture shape and size on observed \lmr. \\

To estimate the uncertainty in \lmr, we used a Monte Carlo method. This involved generating 100 realizations of the galaxy's velocity and velocity dispersion maps using a Gaussian distribution for each Voronoi bin, where the mean is set to the original values and the standard deviation is set to the associated measurement uncertainties. For each simulated set of kinematic maps, we recalculated \lmr\ using \autoref{eq:lmr}. The standard deviation of the resulting \lmr\ distribution was then adopted as the uncertainty on the observed value. The median uncertainty in the \lmr\ for the sample is 0.003 with a sample standard deviation of 0.004.  
\begin{figure*}
    \centering
    {\includegraphics[scale=0.6]{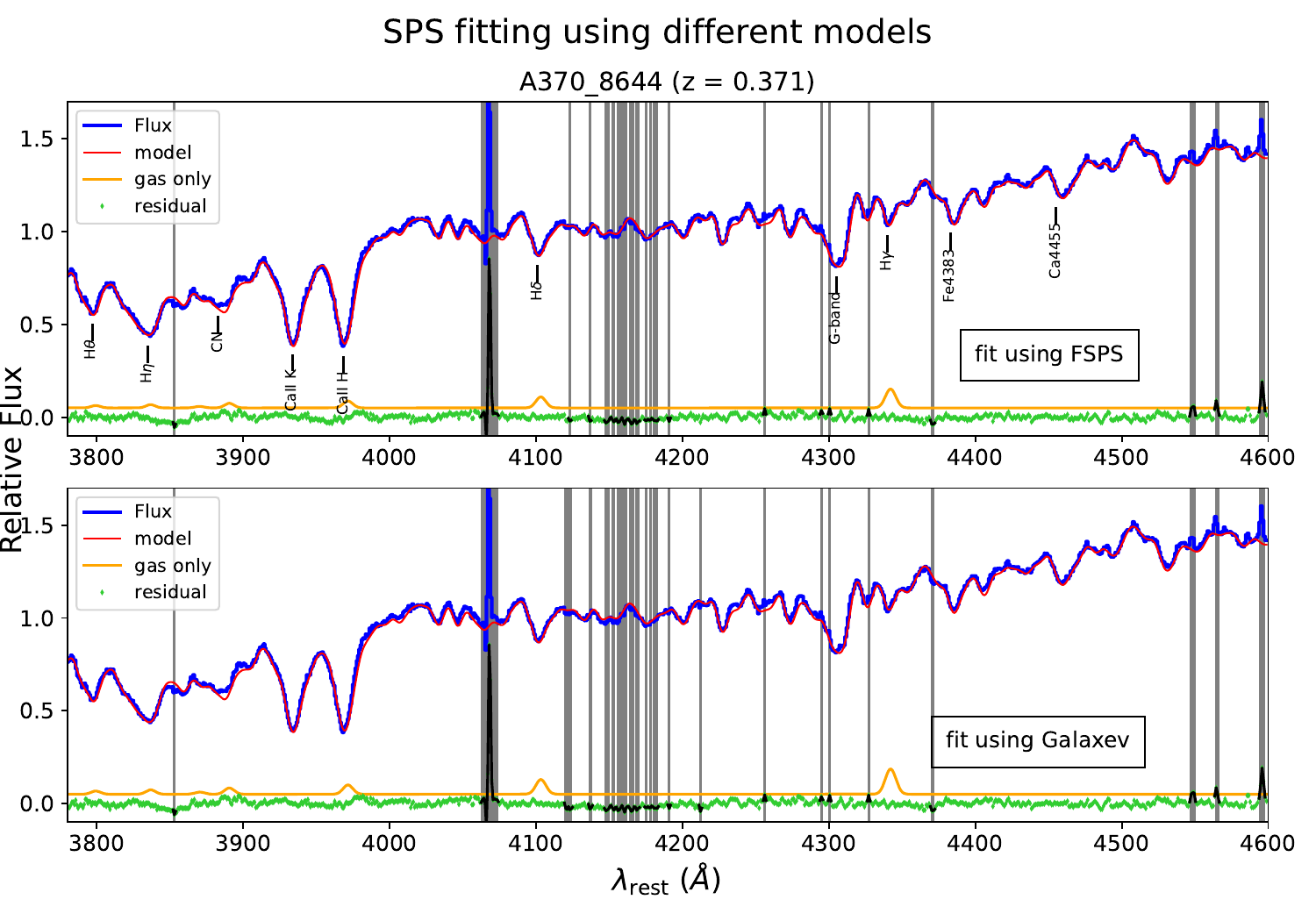}}
    \caption{Example of SPS fitting from \textsc{pPXF}. Top-fitting was done using FSPS templates. Bottom-fitting was performed using GaLAXEV templates. In both cases, the blue curve shows the observed galaxy spectrum from A370\_8644 (z=0.371), the red curve is the best-fitting model (stellar continuum + gas) from \textsc{pPXF}, the orange curve is the best-fitting gas-only spectrum, and the green diamonds are the residuals. The grey-shaded regions were masked and excluded from fitting.}
   \label{fig:sps_fit}
\end{figure*}

\subsubsection{Kinematic major axis}
\label{sec:kin_PA}
We measured the kinematic PA of each galaxy using the median-subtracted 2D velocity maps. For this task, we used the `fit\_kinematic\_pa' function from the $\textsc{PAFIT}$ package. This function implements the algorithm described by \citet[Appendix~C]{pafit_Krajnovic_2006}. Briefly, the routine generates a number of bi(anti)-symmetric velocity maps with different angles. In these models, the velocity of the light-weighted centroid coordinate of a Voronoi bin is set to the weighted average of the data from four quadrants such that
\begin{equation}
     V^{\prime} (x, y) = \frac{V(x, y) + V(x, -y) - V(-x, y) - V(-x, -y)}{4}
\end{equation}
where V$^{\prime}$ is the model velocity field, V is the observed velocity field, and the x-axis is along the chosen angle. The best-fitting position angle, PA$_{\text{kin}}$ is determined by minimizing the $\chi^2$ difference between the model and the observed velocity field. The adopted uncertainty in PA$_{\text{kin}}$ is the formal error provided by the `fit\_kinematic\_pa' function.

\subsection{Measuring stellar population}
\label{sec:sps_prop}
The mass assembly and chemical enrichment histories are a key part of the galaxy evolution process, and stellar population synthesis (SPS) could help decode this information about a galaxy. One way to conduct this stellar population synthesis is to perform a full spectral fitting using templates from Stellar population synthesis (SPS) models. In this section, we discuss how the global stellar population properties were extracted using the integrated spectra within the half-light isophote.

\subsubsection{SPS models}
We used two independent SPS models to check the sensitivity of the results to the adopted model assumptions. These two models were selected because they include templates down to an age of 1 Myr, needed to reproduce actively star-forming galaxies if present in our sample. The SPS models used in this work are : (i) Flexible Stellar Population Synthesis (FSPS)\footnote{\hyperlink{https://github.com/cconroy20/fsps}{https://github.com/cconroy20/fsps}} \citep{FSPS_Conroy_2009, FSPS_Conroy_2010}, and (ii) GaLAXEV\footnote{\hyperlink{http://www.bruzual.org/bc03/}{http://www.bruzual.org/bc03/}} \citep{Galaxv_SPS_model}. \\

%\michele[]{Are these the default pPXF SPS, or did you compute different sets?} 
We used the SPS templates available with the \textsc{pPXF} package. In our adopted standard cosmology, the age of the Universe is $6.9-10.5$ Gyr between $z=0.75$ and 0.25. However, we allowed templates beyond this physical age range to check the ability of the data to retrieve the correct ages of the galaxies. Also, no constraint was imposed on metallicity. Using three different SPS models than the ones used here, KM25 found that the models usually choose the correct set of templates in terms of age and metallicity, regardless of constraints imposed on age and metallicity or not. The setup for the two SPS models is described in \citet{LEGAC_ppxf_Cappellari_2023}. \\

% We used the same age grid for both models, where each element in the grid was logarithmically spaced by 0.1 dex from 1 Myr to 15.85 Gyr, defined as
% % $$
% \lg (\text {Age} / \mathrm{yr})=6,6.1,6.2, \ldots, 10.2
% $$

% (i) Using the FSPS model, we created 387 SPS templates with the above age grid and 9 equally-spaced metallicities $[Z /H]=[-1.75, -1.5, -1.25, -1.0, -0.75, -0.5, -0.25, 0.0, 0.25]$. A Salpeter (\citeyear{Salpeter_1955}) IMF with a mass range of 0.08 - 100 $\mathrm{M}_{\odot}$ was adopted along with the MIST isochrones \citep{MESA_isochrone_Choi_2016}. We computed the SPS spectra \michele[what do you mean? gas masked? and for dust?]{excluding the effect of gas or dust} and adopted default values for the other parameters. This SPS use the MILES stellar library \citep{Sanchez_2006, MILES_sp_model_Falcon_2011} as empirical stellar spectra in the optical region.

% (ii) The GALAXEV model was used to generate a set of 215 template spectra with the age grid mentioned above and 5 metallicities $[Z / H]=[-1.74, -0.73, -0.42, 0.0, 0.47]$. A Salpeter IMF and Padova isochrones \citep{Isochrone_Girardi_2000, Isochrone_Marigo_2008} were used. This model also uses the MILES library to produce the templates at the optical region. 

\subsubsection{Fitting procedure}

To analyze the global stellar population properties of a galaxy, we fitted the SPS templates to its integrated spectrum within the half-light isophote using \textsc{pPXF}. We performed the SPS fitting separately from the kinematic one, as a different setup was used for each. The galaxy spectra and the SPS templates were prepared following the same steps as for kinematic fitting, and we used the same rest-frame wavelength range of 3780-4600 \r{A}. However, unlike kinematic fitting, we simultaneously fitted both the stellar continuum and gas emission lines of the galaxy spectra this time. For this, a set of gas emission lines was included in the template that fall within the adopted rest-frame wavelength range. These emission lines are Balmer series bluer than $\mathrm{H} \beta$, and [NeIII] $\lambda \lambda 3868,69$ doublets. We forced the kinematics of all the gas lines to be the same, and the Balmer series to be fitted as a single gas template. Additionally, we required that the velocity dispersion of the gas emission lines be smaller than the stellar one $\sigma_{\text{gas}} < \sigma_{\text{star}}$. Although this constraint is not expected to be strictly satisfied in all galaxies, and should not be applied if the goal is to study gas kinematics, \citet{LEGAC_ppxf_Cappellari_2023} found that it is necessary to avoid degenerate cases---such as when the spectrum lacks gas emission lines and the Gaussian component intended for emission becomes so broad that it mimics the stellar continuum. The template spectra were normalized using the total bolometric flux within the wavelength range of 1000-30000 \AA. 

The \textsc{pPXF} fitting was performed in two steps. First, we used an iterative sigma-clipping procedure to identify pixels to be masked in the final fit and to rescale the noise spectrum so that the reduced $\chi^2$ equals 1. Next, the final fit was carried out using the rescaled noise spectrum, excluding the previously masked pixels. Unlike the kinematic fitting, we omitted additive polynomials and included only multiplicative polynomials of degree 2, since multiplicative polynomials preserve line indices. Additive polynomials are mainly useful for correcting template mismatch, AGN contamination, or sky subtraction errors, while multiplicative polynomials address issues with spectral flux calibration or reddening. We also applied mild regularization (`regul'=10) to the stellar components to suppress noise in the SPS models (see \citet[fig.~5]{LEGAC_ppxf_Cappellari_2023} for an illustration of regularization effects on SPS fitting). An example of SPS fits to a galaxy spectrum using two different template sets is shown in \autoref{fig:sps_fit}.

%The SPS templates were also log-rebinned at the same velocity scale and smoothed to account for the lower instrumental resolution of the galaxy spectra. The FWHM of the Gaussian kernel used to smooth the templates was set to the difference between the rest-framed FWHM of the galaxy and that of templates, \michele[new]{as described in \autoref{sec:stellar_kinematics}.}

\subsubsection{Estimating stellar population values}\label{sec:measure_spp}
The \textsc{pPXF} fitting for each galaxy provides the weights of the fitted SPS templates over the 2D age-metallicity grid. Considering only the weights of the stellar components in the templates, the luminosity-weighted (as the templates were already normalized by their total bolometric luminosity) population quantities, e.g. age and metallicity, can be calculated by summing the best-fitting template weights such that
\begin{equation*}
    \begin{aligned}
        \langle\lg \text { Age }\rangle & =\frac{\sum_j w_j \times \lg \mathrm{Age}_j}{\sum_j w_j} \\
        \langle[Z / H]\rangle & =\frac{\sum_j w_j \times[Z / H]_j}{\sum_j w_j} 
\end{aligned}
\end{equation*}
where $w_j$ is the weight of the $j$ th template. These equations were implemented using the `sps\_util.mean\_age\_metal' function in \textsc{pPXF}. The stellar mass-to-light ratio is defined as 
\begin{equation*}
    M_* / L=\frac{\sum_j m_j M_{*, j}}{\sum_j m_j L_j} \text {, }
\end{equation*}
where $M_{*, j}$ is the stellar mass, $m_j$ is the mass-weight and $L_j$ is the corresponding luminosity of the $j$th template. %measured in the observed 'F814W' band at the redshift of the galaxy. The mass weights are calculated from the best-fitting template weights. 
This calculation was done by the `mass\_to\_light' function in \textsc{pPXF}. The computation was performed in the HST band used to observe the galaxy and on the redshifted spectrum (using the `redshift' keyword). This $M_* / L$ only includes the mass of living stars and stellar remnants, but excludes the gas lost during stellar evolution. 

To estimate the stellar mass, $M_*$, we first measured the luminosity of the galaxy from the corresponding MGE models in the HST band, accounting for cosmological dimming and extinction. We converted to solar luminosities using the solar luminosity in the same HST band on the redshifted solar spectrum as the flux reference. The solar luminosity was measured using the `mag\_sun' function from \textsc{pPXF}. We then multiplied the galaxy's luminosity by its $M_*/L$, in the same redshifted band, to derive the $M_*$ of the galaxy. 

%See example here at the section "Alternative stellar mass using a single band" \url{https://github.com/micappe/ppxf_examples/blob/main/ppxf_example_legac_photometry.ipynb}]{We have estimated the stellar mass,} $M_{\ast}$, of each galaxy by multiplying the respective $M_* / L$ with the total luminosity of the galaxy measured from the MGE model. The surface brightness of the MGE models was converted to the AB magnitudes and corrected for the extinction effects. The corrected surface brightness was multiplied by $(1 + z)^3$, with z being the galaxy redshift, to account for \michele{both the $(1 + z)^4$ bolometric cosmological dimming and the redshifting of the bandpass in the AB system}.  As the MUSE galaxies are at significant redshift, we also applied K-correction \citep{K-correction_Hogg_2002}. The K-correction relates the observed photometry at a particular bandpass to its intrinsic value at the rest-frame bandpass. For K-correction, we used the absolute solar magnitude of the sun computed in the AB magnitude system with the observed filter ('F814W', 'F606W', etc.) at the redshift of the galaxy. The absolute magnitude of the sun was measured using the 'mag\_sun' function from 'SPS\_UTIL' of pPXF.

\begin{figure*}
    \centering
    \includegraphics[width=\textwidth]{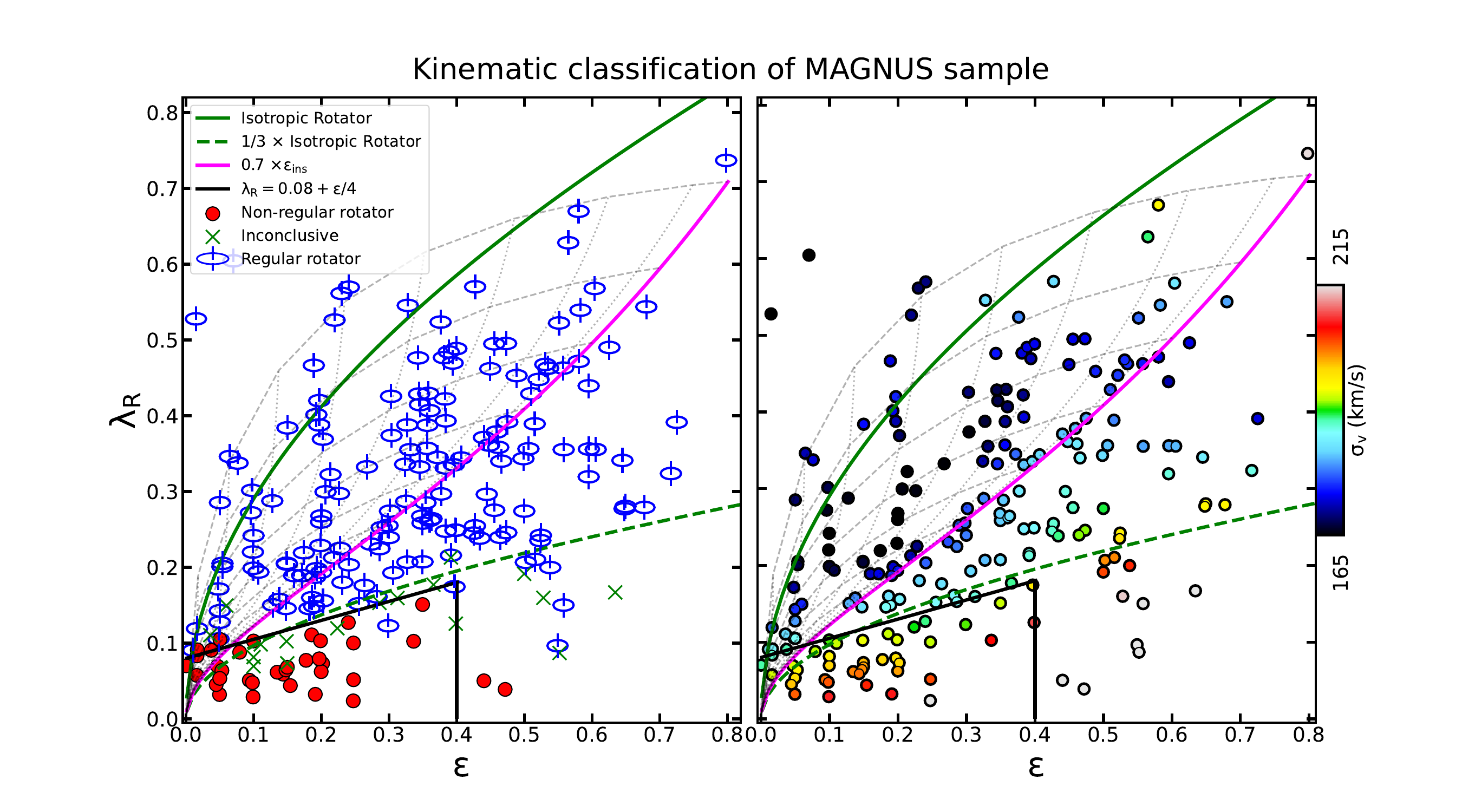}
    \caption{Distribution of $\lambda_R$ in the MAGNUS sample.
    \textbf{Left:} The specific angular momentum proxy $\lambda_R$ is plotted against ellipticity $\epsilon$. $\lambda_R$ was measured within an elliptical aperture with a semi-major axis of 2\remaj, and $\epsilon$ is the observed ellipticity from MGE models. Markers indicate visual classifications: blue disks for regular rotators, red circles for non-regular rotators, and green crosses for unclassified galaxies. The solid green line shows the theoretical relation for an edge-on ($i=90^{\circ}$) isotropic rotator \citep{Binney2005}, using the formula from \citet[][equation 14]{Cappellari_review_2016}, while the dashed green line is at one-third of this value. The magenta line represents the empirical anisotropy limit for edge-on local galaxies \citep[fig.~9]{Cappellari_2007}. Grey dotted lines show this limit for different inclinations (in steps of $\Delta i=10^{\circ}$), and grey dashed lines illustrate how galaxies with given intrinsic ellipticities (in steps of $\Delta\epsilon_{\text{intr}}=0.1$) move on the diagram as inclination changes. The black solid lines, defined by $\lambda_{R_{\mathrm{e}}}<0.08+\varepsilon_e / 4$ and $\varepsilon_e<0.4$ \citep[][equation~19]{Cappellari_review_2016}, separate slow from fast rotators, updating the previous criterion by \citet{ATLAS_III_Emselem_2011} (green dashed line). Our measured \lmr\ values generally confirm the visual classifications.
    \textbf{Right:} Distribution of LOESS-smoothed velocity dispersion $\sigma_\mathrm{e}$ (see \autoref{sec:result_kin_classifiation}) on the ($\lambda_R, \epsilon$) plane. This panel shows that non-regular rotators typically have higher $\sigma_\mathrm{e}$ than most regular rotators. The lines are identical to the left panel.}
    \label{fig:lambda_ep}
\end{figure*}

\begin{figure}
    \centering
    \includegraphics[width=\columnwidth]{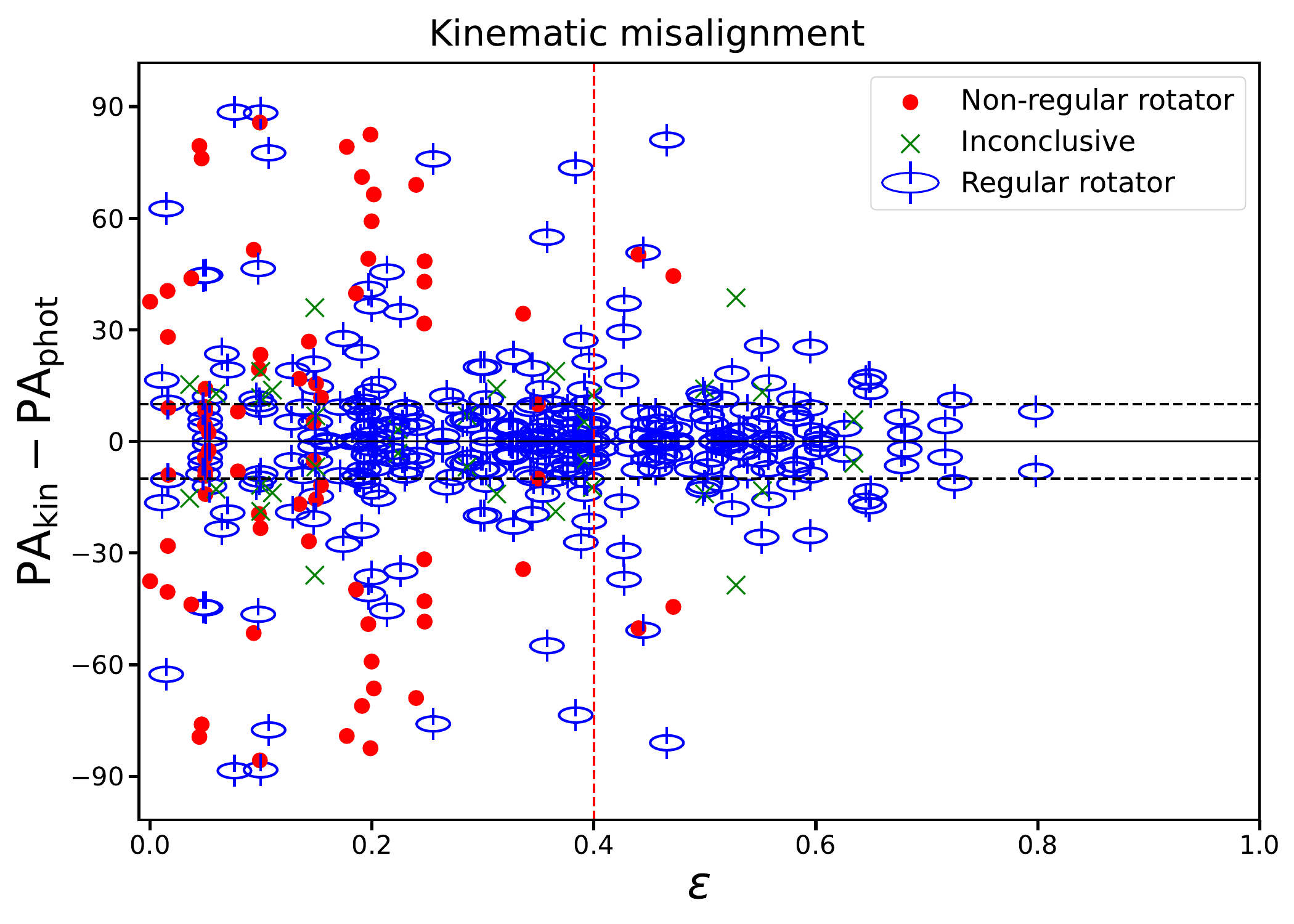}
    \caption{Misalignment angle between kinematic and photometric PAs as a function of $\epsilon$. $PA_{\text{phot}}$ of the galaxies was measured using the sky-subtracted HST images, and $PA_{\text{kin}}$ was measured from the median-subtracted velocity maps. The blue, red, and green markers denote the same qualitative classification as described in \autoref{fig:lambda_ep}. The plot is symmetrized about the black solid line along 0$^{\circ}$ to eliminate any indication of bias towards positive or negative misalignments. The dashed lines show $\pm 10^{\circ}, \pm 30^{\circ}$ misalignments. Most of the regular rotators have $\Psi_{\text{mis}} < |10^{\circ}|$.}
    \label{fig:kin_phot_pa_diff}
\end{figure}

\begin{figure*}
    \centering
    {\includegraphics[width=\textwidth]{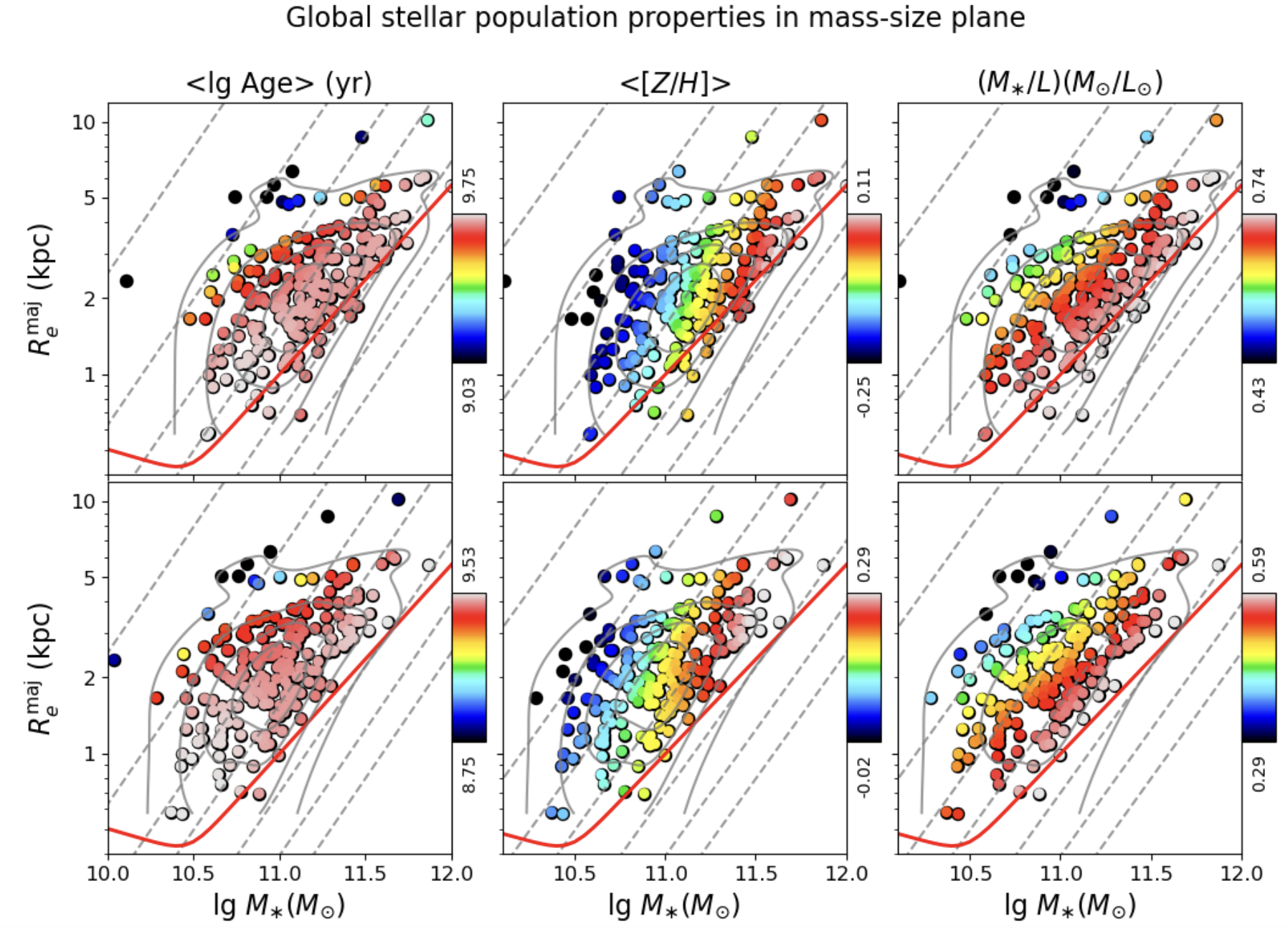}}
    \caption{Age, metallicity and ($M_{\ast}/L$) estimates from SPS models in the mass-size plane. Distributions of luminosity-weighted and LOESS-smoothed age (left panel), metallicity, $\langle[Z/H]\rangle$ (middle panel), and stellar mass-to-light ratio, ($M_{\ast}/L$) (right panel) on the mass–size plane (lg \remaj\ vs lg $M_{\ast}$). The galaxy semi-major axis within the half-light elliptical isophote, \remaj\ was measured from MGE models (see \autoref{sec:measure_structure}), and stellar mass, $M_{\ast}$ was measured by multiplying the estimated $M_{\ast}/L$ by the total luminosity (see \autoref{sec:measure_spp}). The top panel shows results from SPS analysis with FSPS templates and the bottom with  GaLAXEV templates. In each panel, the gray dashed lines indicate the lines of constant $\sigma_\mathrm{e}$ for 50, 100, 200, 300, 400 and 500 km s$^{-1}$ from left to right. 
    %and the black dotted \michele[I would remove the surface density from this plots as it is not critical and the plot looks messy.]{lines are for constant surface mass density, $\Sigma_\mathrm{e} = 10^8, 10^9, 10^{10}, 10^{11}$ M$_{\odot}$ kpc$^{-2}$ from top to bottom.} 
    The red solid curve marks the zone of exclusion (ZOE) from \citet{ATLAS_XX_LOESS_Cappellari_2013b}. A kernel density estimate of the galaxy number density is also indicated by the grey contour lines.}
    \label{fig:age_metal_ml}
\end{figure*}

\begin{figure*}
    \centering
    {\includegraphics[width=0.7\textwidth]{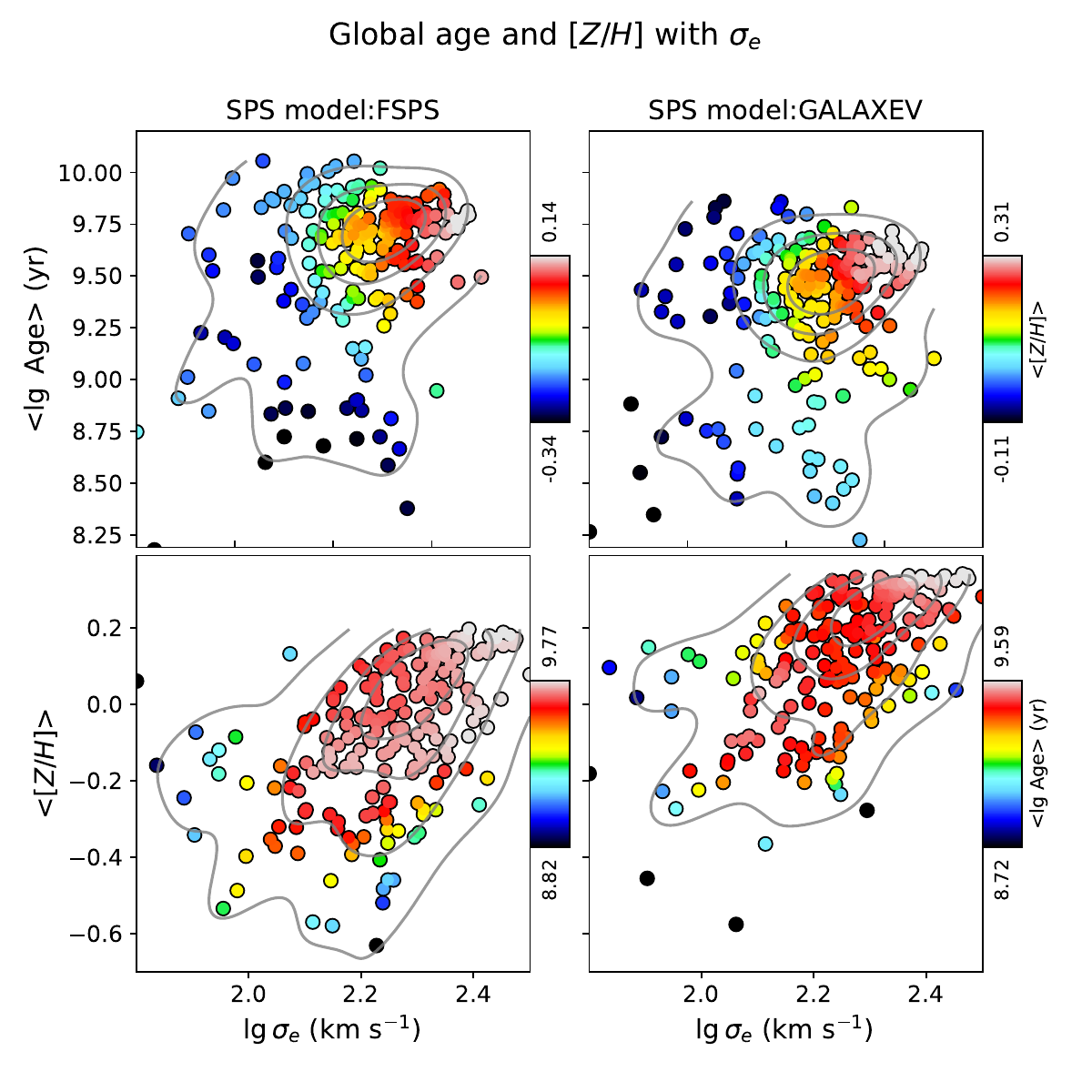}}
    \caption{Luminosity-weighted age ($<\text{lg age}>$) and metallicity , $\langle[Z/H]\rangle$, from different SPS models as a function of $\sigma_\mathrm{e}$. The top panel shows the distribution of age colored by LOESS-smoothed global metallicity, and the bottom panel shows the distribution of the metallicity colored by LOESS-smoothed age. In both rows, the left column shows results from fitting with FSPS templates and the right column from GaLAXEV templates. In each plot, a kernel density estimation of the galaxy number density is indicated by the grey contours.}
    \label{fig:age_metal_vs_sigma}
\end{figure*}

\begin{figure*}
    \centering
    {\includegraphics[width=0.9\textwidth]{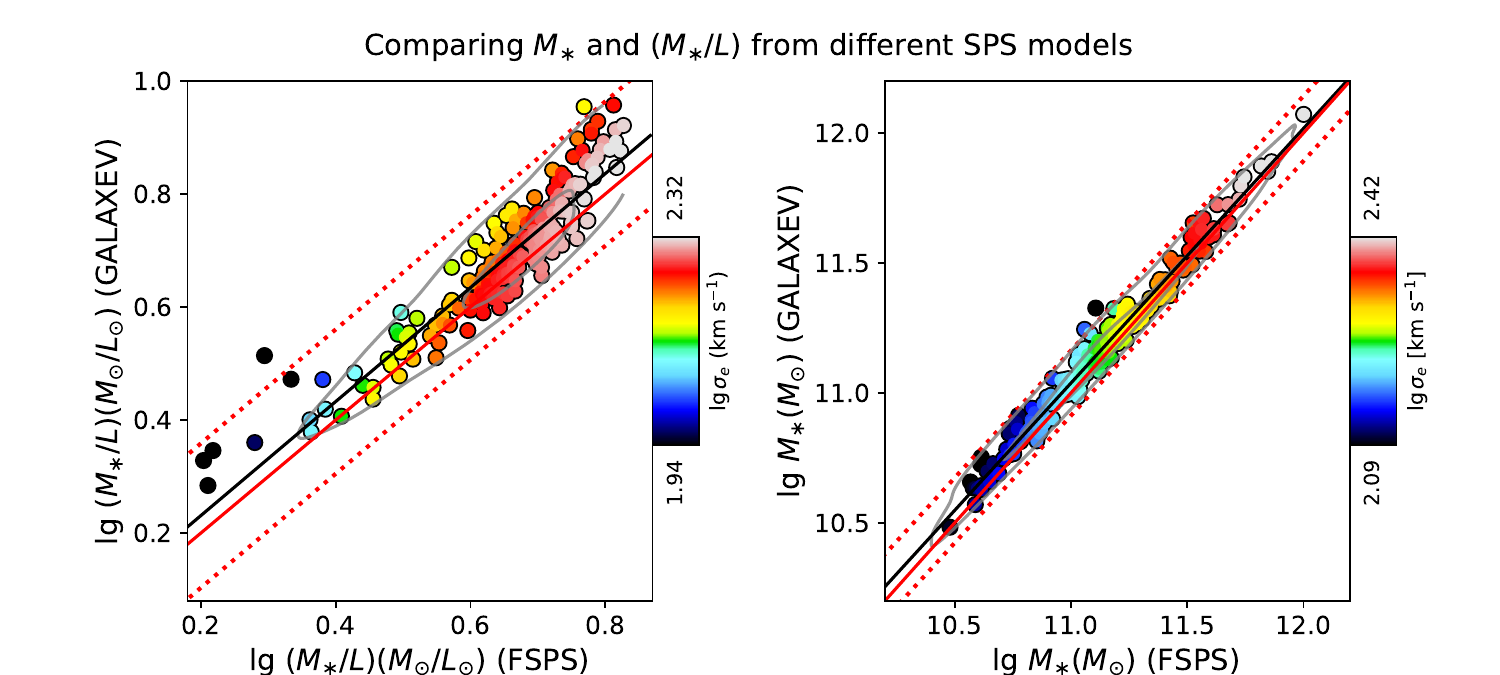}}
    \caption{
    %\michele[]{When comparing identiocal quantities on both axes,I would try to use identical axes ranges, with identical scaling and a line indicating the one-to-one relation.}
    Comparing stellar mass, $M_{\ast}$ (right panel) and mass-to-light ratio, $M_{\ast}/L$ (left panel) from two different SPS models colored by LOESS-smoothed velocity dispersion, $\sigma_\mathrm{e}$. In both plots, the black line denotes the best-fit line, and the red dotted lines show the selection region of galaxies for fitting. The coefficients of the best-fit lines were measured using the \text{LtsFit}python package \citep{ATLAS_XV_Michele_2013}. The red line shows the one-to-one relation between the measurements from different SPS models. A kernel density estimate of the galaxy number density is also indicated by the grey contour lines. In the case of $M_{\ast}/L$, the one-to-one line distorts significantly from the best-fit line. We have to shift the estimates from the Galaxev model, both $M_{\ast}$ and $M_{\ast}/L$, by 0.2 dex to bring them close to the one-to-one relation. As expected, both $M_{\ast}$ and $M_{\ast}/L$ estimates from each model increase with $\sigma_\mathrm{e}$.}
    \label{fig:comp_mass_ml}
\end{figure*}

\section{Results and discussions}
\label{sec:results_discussion}

\subsection{Kinematic classification}\label{sec:result_kin_classifiation}
ETGs could be broadly classified into two distinct groups-- regular (or fast) and non-regular (or slow) rotators \citep{Kinematic_classification_Emsellem_2007, ATLAS_III_Emselem_2011}. Regular rotators exhibit an ordered rotational motion usually extending beyond 1\remaj\ and a well-defined symmetry axis, typically the same as the photometric major axis. Their velocity fields are generally accurately described by that of a thin rotating disk and resemble a ``butterfly pattern" when appropriately color mapped (e.g., A370\_8644 in \autoref{fig:kin_measurement}). In contrast, non-regular rotators do not exhibit an ordered rotational motion or a well-defined symmetry axis and are usually spheroidal and lacking stellar disks. Their velocity fields show more complex and disordered kinematic features (e.g., MACS1149\_62 in \autoref{fig:kin_measurement}). 

%This kinematic classification improves our understanding of galaxy formation pathways, mergers, and assembly history, and the environmental effects on redshift evolution \citep{Cappellari_review_2016, Munoz_Lopez_2024}.
In \autoref{fig:lambda_ep}, we plotted the observed $\lmr$ as a function of $\epsilon$ for the MAGNUS sample and marked the galaxies using their visual classification (regular, non-regular rotators, and inconclusive class). Regular rotators are typically observed to exhibit higher specific angular momentum (higher \lmr), while non-regular rotators tend to have lower values, both in the local Universe \citep[e.g.][]{ATLAS_III_Emselem_2011, Vande_sande_SAMI_2017, Graham_2018, Dynpop_I_Zhu_2023} and at intermediate redshifts, where this dichotomy was first established using long-slit spectroscopy \citep[e.g.][]{vander_Marel_2007, vandaer_Wel_2008, LEGAC_Bezanson_2018, van_Houdt_2021} and subsequently confirmed with IFU studies \citep[e.g.][]{Munoz_Lopez_2024, Derkenne_2024}.  We observed a similar trend for the MAGNUS sample. This suggests that the kinematic state of ETGs has remained broadly the same below z $\sim$ 1. More broadly, kinematic studies of quiescent galaxies have 
now been extended to $z \sim 1.5$--$3$ through a combination of deep spectroscopy, gravitational lensing, and space-based observations \citep[e.g.][]{Belli_2017, Toft_2017, Newman_2018, DEugenio_2024}, demonstrating that quiescent galaxies at these epochs are predominantly fast-rotating disk-like systems with substantial rotational support. \citet{Cappellari_review_2016} proposed the black solid line defined by $\left(\lambda_{R_{\mathrm{e}}}<0.08+\varepsilon_e / 4, \varepsilon_e<0.4\right)$ to qualitatively separate slow rotators (below these lines) from the fast rotators. This updates the criterion by \citet[green dashed line]{ATLAS_III_Emselem_2011} with the introduction of a limit on ellipticity, to reduce the contamination from counter-rotating disks. In general, our qualitative classification of regular and non-regular rotators is in concordance with these two classification schemes. Around 87\% of galaxies that were qualitatively classified as non-regular rotators reside within the slow rotator region bordered by the black solid lines, while 99\% of galaxies that were visually classified as regular rotators stay within the fast rotator regime (outside the black solid lines). From \autoref{fig:lambda_ep} it is evident that the classification scheme proposed by \citet{Cappellari_review_2016} works relatively better for our sample than the one from \citet{ATLAS_III_Emselem_2011}. Thus, we adopted the first one as the quantitative classifier of the kinematic state in MAGNUS galaxies. We note that the kinematic classification presented here is based on observed \lmr\ values without PSF correction, which causes them to be systematically lower than the intrinsic value. The impact of PSF correction on the observed \lmr\ measurements of the MAGNUS sample is discussed in detail by \citet{Magnus_II} (MAGNUS~II). \\

According to quantitative classification, the slow rotator fraction in our sample is 19.3$^{+3.0}_{-2.4}$\% (41 out of 213). The uncertainty was measured using a beta distribution quantile technique recommended by \citet{Ewan_2011_bionomial_uncertainty} to estimate the uncertainty of a binomial distribution. Using the same quantitative classification \citet{Derkenne_2024} found that the slow rotator fraction in MAGPI sample (z $\sim$ 0.3) is 18$^{+6.6}_{-4.1}$\%\% with seeing corrected \lmr\ and is 24$^{+6.9}_{-4.9}$\% without seeing corrected \lmr. The MAGPI sample is mass-matched to our sample, as in both samples the estimated stellar masses of the galaxies are $M_{\ast} > 3 \times 10^{10} M_{\odot}$. They conducted a comparison with the local Universe ETGs by drawing a mass-matched and progenitor bias-corrected sample from the MaNGA survey \citep{Freedman_2020} 100 times. They found that the slow rotator fraction in MaNGA is 19$\pm$6\% and suggested that the slow rotator population of ETGs has not changed significantly since z $\sim 0.3$. Comparing the LEGA-C ETG sample with local counterparts, \citet{Cole_2020} also found that the slow rotator fraction at $z \sim 0.8$ is broadly consistent with the local Universe up to $\log M_{\ast}/M_{\odot} \sim 11.5$. The slow rotator fraction in the MAGNUS sample also hints towards this same picture. However, \citet{Munoz_Lopez_2024} found almost no slow rotators in a sample of 106 galaxies with a redshift range of 0.1 $< z <$ 0.8. This is probably because their sample only contain 5\% galaxies with stellar mass, $M_{\ast} > 10^{11} M_{\odot}$ while 85\% of the slow rotators in our sample have $M_{\ast} > 10^{11} M_{\odot}$. The slow rotator population in the local Universe is also dominated by galaxies with $M_{\ast} > 10^{11} M_{\odot}$ \citep{Cappellari_review_2016}. Note that for the MAGNUS sample, we have used the stellar mass measurements from the FSPS templates. We found no significant change in the slow rotator fraction over redshift within the MAGNUS sample.
%The MAGNUS sample, spanning $0.25 < z < 0.75$ with homogeneous IFU data, provides a critical link between these high-redshift kinematic studies and the well-characterized local ETG population.

\autoref{fig:lambda_ep} also shows the distribution of the MAGNUS galaxies in the ($\lmr$, $\epsilon$) plane, colored by their LOESS-smoothed velocity dispersions, $\sigma_\mathrm{e}$. The 2D `Locally Weighted Regression' (LOESS) algorithm \citep{LOESS_algorithm_Cleveland1988}, implemented in the LOESS package\footnote{\url{https://pypi.org/project/loess/}} \citep{ATLAS_XX_LOESS_Cappellari_2013b}, is the two-dimensional equivalent of plotting a mean or median trend line through a one-dimensional scatter plot. While in one dimension it is common to visualize the scatter of the data around the mean trend, this is not easily achievable in two dimensions. The LOESS visualization intentionally smooths over the scatter in the data to reveal the underlying average trend, which would otherwise be difficult to see. We used a smoothing parameter `frac` = 0.3. As expected, the smoothed map shows that slow rotators generally have higher $\sigma_{\mathrm{e}}$ than fast rotators, consistent with findings in the local Universe \citep[e.g.][]{Graham_2018} and at intermediate redshifts \citep{LEGAC_Bezanson_2018, Cole_2020}. We found that around 80\% of the slow rotators in this sample have $\sigma_\mathrm{e} >$ 175 km s$^{-1}$ while only 50\% of the fast rotators have such high $\sigma_\mathrm{e}$.

\subsection{Kinematic misalignment}
\label{sec:kin_misalignment}
The kinematic misalignment angle, $\Psi_{\rm{mis}}$, the angular separation between the measured photometric major axis and the kinematic major axis, is then defined as
\begin{equation*}
    \Psi_{\rm{mis}} = \text{PA}_\text{kin} - \text{PA}_\text{phot}
\end{equation*} 
The misalignment between the kinematic and photometric position angles ($\Psi_{\rm{mis}}$) for the MAGNUS sample is presented in \autoref{fig:kin_phot_pa_diff}, which is symmetrized around $\Psi_{\rm{mis}} = 0^{\circ}$ to avoid any bias toward positive or negative values. The uncertainty in this measurement is dominated by the error in the kinematic position angle, $\text{PA}_\text{kin}$. As expected, the median uncertainty in $\text{PA}_\text{kin}$ is significantly higher for non-regular rotators ($\sim 30^{\circ}$) than for regular rotators ($7^{\circ}$), because the former lack an ordered velocity field and thus a well-defined kinematic major axis.

We find that approximately 60.0\% of the quantitatively confirmed regular rotators exhibit a small misalignment with $\Psi_{\rm{mis}} < |10^{\circ}|$. This finding is consistent with observations of low-redshift regular rotators in numerous studies \citep[e.g.,][]{Cappellari_2007,Krajnovic2011,Cappellari_review_2016, Graham_2018}. For comparison, a study of 2,286 galaxies from the MaNGA survey by \citet{Graham_2018} found that 70.0\% of regular rotator ETGs and 56.3\% of spiral galaxies have kinematic misalignments within $\pm 10^{\circ}$. That study also noted that apparent misalignment increases with lower data quality, suggesting that measurement uncertainties are a primary source of scatter. When we apply a similar quality cut to our sample, considering only the regular rotators with a $\text{PA}_\text{kin}$ uncertainty below the median of $7^{\circ}$ (82 out of 172), the fraction with $\Psi_{\rm{mis}} < |10^{\circ}|$ increases to 73\%. 
%%This strong agreement, despite differences in distance, data quality, and sample size, suggests that our findings are consistent with local Universe results and indicates that the kinematic structure of regular rotators has not changed dramatically over the last 5-6 billion years.
These results represent the most statistically significant IFU-based measurement of stellar kinematic--photometric alignment for massive quiescent ETGs at $0.25 < z < 0.75$ to date, complementing the earlier MUSE-based study of \citet{MUSE_HUDF_kinematics_Adrien_2017} which was limited to 17 galaxies at $0.2 \lesssim z \lesssim 0.8$. The close alignment between the kinematic and photometric axes of fast rotators across this redshift range indicates that the internal dynamical structure of massive ETGs has remained stable over the last $\sim$7~Gyr.
%Given our small sample size, it is hard to evaluate whether this discrepancy happened due to poor accuracy, as we are dealing with objects from comparably larger distances, or it's a feature of high-redshift regular galaxies. 
%However, 90\% of the qualitatively confirmed regular rotators have misalignment angle $\Psi < |30^{\circ}|$.

\subsection{Global stellar population properties}
\label{sec:stellar_pop}
\autoref{fig:age_metal_ml} presents the LOESS-smoothed distributions of global stellar population properties --- age, metallicity ($\langle[\mathrm{Z/H}]\rangle$), and stellar mass-to-light ratio ($M_{\ast}/L$) --- on the mass--size ($M_{\ast}$, \remaj) plane, derived using two different SPS models. The semi-major axis \remaj\ was measured from MGE models (\autoref{sec:measure_structure}), while the stellar mass $M_{\ast}$ was derived from the $M_{\ast}/L$ and total luminosity (\autoref{sec:measure_spp}). We also plot the `zone of exclusion' (ZOE) for nearby galaxies from \citet{ATLAS_XX_LOESS_Cappellari_2013b}, but scaled it down by a factor of 1.6 in \remaj, roughly consistent with the general trends of decreasing galaxy sizes with redshift \citep[e.g.][]{vanderwel_2014}. The ZOE is a structural lower limit on galaxies below which the combination of high mass and compact size would imply unphysically high velocity dispersions or densities, constrained by galaxy formation physics and the virial theorem. \autoref{fig:age_metal_vs_sigma} shows how luminosity-weighted age and metallicity vary with $\sigma_{\mathrm e}$. The following subsections discuss each stellar population property in turn, comparing our results with local Universe studies and prior work at intermediate and high redshifts. In \autoref{sec:kin_pop}, we briefly examine how these properties depend on the kinematic state of the galaxies, as quantified by \lmr.

\subsubsection{Age--$\sigma_{\mathrm{e}}$ Relation}
\label{sec:age_sigma}
In \autoref{fig:age_metal_ml}, stellar ages of the MAGNUS sample on the mass--size plane show a general trend consistent with lines of constant $\sigma_{\mathrm{e}}$, with the most massive and compact galaxies tending to be the oldest. At fixed $M_{\ast}$, larger galaxies are on average younger than their more compact counterparts, consistent with the picture that compact galaxies have older stellar populations. This age--size anti-correlation is visible for both SPS models, though with considerable scatter.

\autoref{fig:age_metal_vs_sigma} (top panels) shows the distribution of luminosity-weighted age as a function of $\sigma_{\mathrm{e}}$, colored by LOESS-smoothed metallicity. The overall dependence of age on $\sigma_{\mathrm{e}}$ is not strongly pronounced, likely because our ETG selection results in a sample with a narrow range of old stellar ages, making any underlying trend difficult to detect against measurement uncertainties. Nonetheless, at any given $\sigma_{\mathrm{e}}$, while the scatter in age is large, older galaxies tend to be more metal-rich than their younger counterparts, as also visible in the bottom panels of \autoref{fig:age_metal_vs_sigma}, where metallicity is colored by LOESS-smoothed age. The recovered ages for both models are within the limit of the Universe age at the redshift of the sample (around 7~Gyr at $z \sim 0.75$). In general, results from both models are consistent, with \textsc{GaLAXEV} recovering slightly younger ages and higher metallicities than \textsc{FSPS}.\\

The correlations we observe between stellar age and $\sigma_{\mathrm{e}}$ in the MAGNUS sample, that ETGs with higher velocity dispersions formed their stars relatively earlier than less massive galaxies, are consistent with the well-established picture of archaeological downsizing \citep[e.g.,][]{Cowie_1996, Thomas_2005, Gallazzi_2005}. Other observed trends in this work between age-size-mass-metallicity are also consistent with a substantial body of prior work at both the local Universe and at intermediate and high redshifts (z$>$1). In the local Universe, the age--$\sigma_{\mathrm{e}}$ relation for ETGs is well established \citep[e.g.,][]{Thomas_2005, Gallazzi_2005, ATLAS_XX_LOESS_Cappellari_2013b, McDermid_2015, Scott2017, Li2018_MaNGA, DynPop_II_Lu_2023}, with more massive, higher-$\sigma_{\mathrm{e}}$ galaxies hosting older stellar populations on average. However, this trend is typically weak or flat relative to the metallicity-$\sigma_{\mathrm{e}}$ relation, when the sample is restricted to ETGs alone, especially in the absence of low-$\sigma_{\mathrm{e}}$ galaxies, consistent with what we observe in the MAGNUS sample.\\

At intermediate redshift, \citet{Gallazzi_2014} first characterized the age--mass relation for a sample of $\sim$70 galaxies at $0.65 < z < 0.75$ using deep IMACS/Magellan spectroscopy, and observed that the downsizing trend is already in place at $z = 0.7$. Using stacked spectra of quiescent galaxies over $0.1 < z < 0.7$, \citet{Choi_2014} also found that the age--mass relation is already present over the full redshift range covered by the MAGNUS sample. Subsequent studies using the LEGA-C survey at $z \sim 0.6$--$1.0$ have provided more detailed characterization of these trends with larger samples. Using LEGA-C data \citet{Chauke_2018} demonstrated that velocity dispersion is a better predictor of age than stellar mass alone, finding that higher-$\sigma_{\mathrm{e}}$ galaxies formed their stars earlier and more rapidly. \citet{Barone_2022} found that the age--$\sigma_{\mathrm{e}}$ relation is weaker at intermediate redshift than local Universe, with the age trend shifting from a dependence on surface density ($M/R_{\mathrm{e}}^2$) at low redshift to a milder dependence on $M/R_{\mathrm{e}}$ at intermediate redshift, suggesting that the age--$\sigma_{\mathrm{e}}$ relation is built up over time. Consistent with this, \citet{Kaushal_2024} found no clear age trends with mass or $\sigma_{\mathrm{e}}$ among quiescent LEGA-C galaxies, and \citet{Nersesian_2025} found a flat age--$\sigma_{\mathrm{e}}$ relation for quiescent galaxies, reflecting that most quiescent galaxies are uniformly old regardless of their velocity dispersion. Similarly, \citet{LEGAC_ppxf_Cappellari_2023} and \citet{Gallazzi_2026a} both found that age tracks $\sigma_{\mathrm{e}}$ in the LEGA-C survey. \\

At higher redshifts, \citet{Kirek_2009} and \citet{van_de_sande_2013} provided some of the first direct measurements of stellar population properties for high-redshift early-type galaxies, establishing that massive quiescent systems at $z \sim 2$ are already old and evolved. \citet{Belli_2015} found that older quiescent galaxies tend to be more compact at $1 < z < 1.6$, consistent with the age--size anti-correlation we observe. \citet{Carnall_2019} found a steep downsizing signal of 1.48~Gyr per decade in mass at $1.0 < z < 1.3$. Recently, \citet{Beverage_2025} found no clear trends between $\sigma_{\mathrm{e}}$ and formation time at $z > 1$, though higher-redshift galaxies formed earlier as expected from passive evolution. Together, these studies suggest that while the age--$\sigma_{\mathrm{e}}$ correlation is not yet strongly established at $z > 1$, it becomes progressively more evident toward lower redshifts, consistent with the weak but present trend in the MAGNUS sample at $0.25 < z < 0.75$.

\subsubsection{Metallicity--$\sigma_{\mathrm{e}}$ Relation}
\label{sec:metal_sigma}
The metallicity--$\sigma_{\mathrm{e}}$ trend is the most pronounced of the three stellar population properties we measure. In \autoref{fig:age_metal_ml} (middle panels), $\langle[\mathrm{Z/H}]\rangle$ clearly increases with $\sigma_{\mathrm{e}}$ for both SPS models across the full range of velocity dispersions in our sample, with minimal dependence on mass or size independently. At fixed $M_{\ast}$, smaller galaxies are more metal-rich than their larger counterparts, and conversely, at fixed size, more massive galaxies are more metal-rich. This pattern is consistent with the gravitational potential depth, which is better captured by $\sigma_{\mathrm{e}}$ or $M_{\ast}/R_{\mathrm{e}}$ than by stellar mass or size alone, being the primary regulator of chemical enrichment. Galaxies with deeper potential wells were more efficient at retaining metal-rich gas against supernova or AGN-driven outflows during their star-forming phase, leading to a higher degree of chemical enrichment that is preserved in their stellar populations as a fossil record of their formation history \citep[e.g.,][]{Franx_1990, Thomas_2005, Barone_2022}. \\

\autoref{fig:age_metal_vs_sigma} (bottom panels) further illustrates this picture. At fixed $\sigma_{\mathrm{e}}$, older galaxies tend to be more metal-rich than their younger counterparts. Conversely, at fixed age, galaxies with higher $\sigma_{\mathrm{e}}$ are systematically more metal-rich, reinforcing the primary role of the gravitational potential in driving chemical enrichment. As noted in \autoref{sec:age_sigma}, results from both SPS models are broadly consistent. \\

Unlike the age--$\sigma_{\mathrm{e}}$ relation, which is comparatively weak among ETGs, the strong metallicity--$\sigma_{\mathrm{e}}$ correlation has been a robust feature of stellar population studies since the some of the earliest absorption-line work on local galaxies  \citep[e.g.,][]{Thomas_2005, Gallazzi_2005}, and remains strong in large IFU surveys such as MaNGA \citep{ATLAS_XX_LOESS_Cappellari_2013b, McDermid_2015, Scott2017, Li2018_MaNGA, DynPop_II_Lu_2023}. Across these studies, higher-$\sigma$ ETGs are consistently more metal-rich, pointing to gravitational potential depth as the primary driver of this relation. \\

A series of LEGA-C-based studies has mapped out how this relation behaves at $z \sim 0.6$--$1.0$. \citet{Gallazzi_2014} showed that the metallicity--mass relation is already in place at $z \sim 0.7$, with more massive galaxies being more metal-rich. \citet{Choi_2014} confirmed that elemental abundances remain stable at fixed stellar mass over $\sim$7~Gyr, implying that chemical enrichment was largely complete by this epoch. Placing elemental abundances on the mass--size plane, \citet{Beverage_2021} found that more compact galaxies are more metal-rich at fixed $M_{\ast}$, in agreement with what we observe here. \citet{Barone_2022} demonstrated that the metallicity--($M_{\ast}/R_{\mathrm{e}}$) relation exists at both $z \sim 0$ and $z \sim 0.7$, supporting the theory that the gravitational potential depth regulates stellar metallicity by determining the escape velocity required for metal-rich gas to be expelled before it can be recycled into later generations of stars. With the LEGA-C DR3 sample, \citet{Beverage_2023} confirmed a positive correlation between total metallicity and $\sigma_v$ across 11 elements, with the highest-$\sigma$ galaxies being the most metal-rich. \citet{Nersesian_2025} reported a weak but positive metallicity--$\sigma$ relation among quiescent galaxies, which are more metal-rich than star-forming galaxies at fixed $\sigma$. As with the age--$\sigma_{\mathrm{e}}$ relation, \citet{LEGAC_ppxf_Cappellari_2023} and \citet{Gallazzi_2026a} also recovered a clear metallicity--$\sigma_{\mathrm{e}}$ trend in LEGA-C, while the latter further showed that metallicity increases with both $M_{\ast}$ and $\sigma$, indicating that more massive galaxies have been more efficient at reaching a higher degree of chemical enrichment. Finally, \citet{Hamadouche_2022} studied the evolution of quiescent galaxy properties between $z \sim 1.1$ and $z \sim 0.7$, finding results broadly consistent with the metallicity trends observed here. \\

The chemical enrichment history implied by this relation can be traced to still higher redshifts. \citet{Kriek_2019} measured stellar metallicities and elemental abundance ratios for massive quiescent galaxies at $z \sim 1.4$ and found that the metallicity--$\sigma_{\mathrm{e}}$ relation is already in place by this epoch. More recently, \citet{Beverage_2025} pushed these measurements past $z > 1$ but found no clear $\sigma_{\mathrm{e}}$--metallicity trend there, though the overall picture is consistent with chemical enrichment being regulated by the gravitational potential across cosmic time. Taken together, this body of work indicates that the strong metallicity--$\sigma_{\mathrm{e}}$ relation observed in the MAGNUS sample at $0.25 < z < 0.75$ is a robust and stable feature of the ETG population that is already established by $z \sim 1$ and essentially remains unchanged since.

\subsubsection{$M_{\ast}/L$--$\sigma_{\mathrm{e}}$ Relation}
\label{sec:ml_sigma}
The stellar mass-to-light ratio, $M_{\ast}/L$, shows a somewhat stronger dependence on $\sigma_{\mathrm{e}}$ than age, but less pronounced than metallicity when considered independently. In \autoref{fig:age_metal_ml} (right panels), $M_{\ast}/L$ clearly increases with $\sigma_{\mathrm{e}}$ for both SPS models. At fixed $M_{\ast}$, more compact galaxies have higher $M_{\ast}/L$, and at fixed size, more massive galaxies have higher $M_{\ast}/L$ --- mirroring the trends seen for both age and metallicity. This is expected, as $M_{\ast}/L$ is a combination of both stellar age and metallicity of the underlying stellar population, and suggests that dark matter is subdominant within the effective radius of these galaxies.

In \autoref{fig:comp_mass_ml}, we compare $M_{\ast}/L$ (left panel) and $M_{\ast}$ (right panel) from the two SPS models, color-coded by LOESS-smoothed $\sigma_{\mathrm{e}}$. As expected, both $M_{\ast}/L$ and $M_{\ast}$ from each model increase with $\sigma_{\mathrm{e}}$. The stellar mass estimates appear to show a tighter correlation than the mass-to-light ratios; however, this is a visual artifact, as the intrinsic scatter is identical for both quantities. This is because for each galaxy, the stellar mass is derived as the product of the mass-to-light ratio and the galaxy luminosity, which is identical for both SPS models. The mass estimates from the \textsc{GaLAXEV} model are approximately 1.5 times lower (0.2~dex in log scale) than the corresponding \textsc{FSPS} estimates. We have not investigated the source of this systematic offset, as it is outside the scope of this paper, but it highlights the importance of cross-validating results using different SPS models. Further studies will be required to understand the origin of this discrepancy between the two models. \\

The $M_{\ast}/L$--$\sigma_{\mathrm{e}}$ relation is well established in the local Universe. \citet{ATLAS_XV_Michele_2013} derived total $M/L$ ratios for 260 ATLAS$^{\rm 3D}$ ETGs using dynamical models, and found a tight relation of the form $(M/L) \propto \sigma_{\mathrm{e}}^{0.8}$ with an intrinsic scatter of $\sim$13\%. \citet{ATLAS_XX_LOESS_Cappellari_2013b} further demonstrated that $\sigma_{\mathrm{e}}$ is the primary driver of $M_{\ast}/L$ variations among local ETGs, with stellar population properties following lines of constant $\sigma_{\mathrm{e}}$ in the mass--size plane. These results have been extended to larger samples using the MaNGA survey, where \citet{DynPop_II_Lu_2023} and \citet{Dynpop_III_Zhu_2024} confirmed the tight $M_{\ast}/L$--$\sigma_{\mathrm{e}}$ relation across both ETGs and late-type galaxies. At intermediate redshifts, \citet{LEGAC_ppxf_Cappellari_2023} demonstrated that $M_{\ast}/L$ tracks $\sigma_{\mathrm{e}}$ for LEGA-C galaxies with similar quality to local surveys, while \citet{van_Houdt_2021} confirmed that dynamical $M/L$ ratios are consistent with stellar population-based estimates for quiescent galaxies at this epoch, with dark matter subdominant within $R_{\mathrm{e}}$. The $M_{\ast}/L$--$\sigma_{\mathrm{e}}$ relation we observe in the MAGNUS sample at $0.25 < z < 0.75$ is consistent with these findings, extending the LEGA-C results to lower redshifts and confirming that the relation is in place across the full redshift range probed by our survey.

\subsubsection{Dependence on Kinematic State}
\label{sec:kin_pop}
Although velocity dispersion captures the key trends in stellar population variation, kinematic classification into slow and fast rotators offers an additional dimension for understanding these properties. To examine the dependence of global stellar population properties on \lmr, we compared a subsample of slow rotator ETGs to their fast rotator counterparts within the MAGNUS sample. To minimize the influence of velocity dispersion, we limited both subsamples to ETGs with $\sigma_\mathrm{e} >$ 200 \kmps. This selection yields 58 fast rotators (34\% of the fast rotator population) and 22 slow rotators (54\% of the slow rotator population). A two-sample Kolmogorov–Smirnov (K-S) test confirms that these two groups are statistically distinct in terms of \lmr\ (p-value 2$\times 10^{-15}$), yet statistically consistent in $\sigma_\mathrm{e}$ (p-value 0.3), with median $\sigma_{\mathrm{e}}$ values of $228 \pm 7$ and $224 \pm 4$ km s$^{-1}$ for fast and slow rotators respectively, confirming that the two subsamples are well matched in velocity dispersion. The uncertainties in this and subsequent median values were estimated using a bootstrap resampling method. \\

We found that the median $\langle[Z/H]\rangle$ of the fast and slow rotator groups is 0.04$\pm$0.03 and 0.12$\pm$0.02, respectively. Similarly, the median values of the $M_{\ast}/L$ of these two groups are 5.0$\pm$0.1 and 5.31$\pm$0.15. The median stellar age of slow rotators ($\langle \lg \mathrm{Age} \rangle = 9.78\pm0.02$) is marginally higher than that of fast rotators ($9.74\pm0.02$), though the difference is not statistically significant. These results indicate that, while both subsets exhibit comparable velocity dispersions, slow rotators have higher metallicities and mass-to-light ratios, though no significant difference is observed in their median stellar ages. It thus appears that stellar population properties in ETGs depend on their kinematics state in addition to velocity dispersion. We note that these qualitative trends hold for the full MAGNUS sample, where the median $\langle[\mathrm{Z/H}]\rangle$ for fast and slow rotators is $-0.08\pm0.02$ and $0.05 \pm 0.02$ respectively, the median $M_{\ast}/L$ is $4.64 \pm 0.12$ and $4.94 \pm 0.19$ respectively and almost no difference in median age. The overall decrease in median metallicity and $M_{\ast}/L$ for both groups compared to the $\sigma_{\mathrm{e}}$-matched subsample is expected, as these properties decrease with $\sigma_{\mathrm{e}}$ in general.

\citet{McDermid_2015} found no significant age difference between slow and fast rotators in ATLAS$^{\rm 3D}$, but found fast rotators to be more metal-rich which is the opposite of what we observe. This difference may partly reflect the different kinematic classification schemes adopted, which can yield different class memberships particularly for galaxies with complex or low-amplitude velocity fields. Using the MaNGA survey \citet{DynPop_II_Lu_2023} found that stellar age, metallicity, and $M_{\ast}/L$ all decrease with increasing galaxy rotation (higher \lmr), consistent with our findings, though their sample includes both early and late-type galaxies. While the median stellar ages of the $\sigma_{\mathrm{e}}$-matched fast and slow rotator subsamples are consistent within the uncertainties, the marginal difference we observe is in the same direction as the age--spin anti-correlation found at both local and intermediate redshifts \citep{van_de_sande_2018, Croom_2024, Foster_2025}, though our data do not allow us to confirm this trend. \\

In summary, the stellar population properties of the MAGNUS sample follow a clear hierarchy: metallicity shows the strongest dependence on $\sigma_{\mathrm{e}}$, followed by $M_{\ast}/L$, while age shows only a weak trend. These results confirm that the gravitational potential well depth, as traced by $\sigma_{\mathrm{e}}$, is the primary regulator of stellar population properties in massive ETGs, consistent with both local Universe IFU surveys and prior intermediate-redshift and high redshift studies, indicating that these scaling relations were already in place by $z \sim 0.75$ and have remained stable over the last $\sim$7~Gyr. Beyond velocity dispersion, the dependence of metallicity and $M_{\ast}/L$ on kinematic state suggests that \lmr\ provides an additional dimension for characterizing the stellar population properties of massive ETGs.

\section{Conclusion}
\label{sec:conclusion}
The assembly history of mass and angular momentum in ETGs remains an open area of research. While large local IFU surveys have provided comprehensive insights into the structural, kinematic, and stellar population properties of ETGs, extending these studies to intermediate redshifts with comparable data quality and sample size has remained challenging. To address this, we constructed a homogeneous sample of 213 ETGs at redshifts $0.25 < z < 0.75$ using MUSE-DEEP datacubes, combined with high-resolution HST imaging. Following the KM25 methodology, we measured integrated and spatially resolved stellar kinematics using three clean stellar template libraries (Indo-US, MILES, and XSL), and modeled surface brightness profiles from HST imaging. This combination of high-quality photometric and kinematic data enabled the measurement of various morphological, kinematic, and stellar population properties across the full redshift range of the sample. \\

The key findings of this work are summarized as follows:

\begin{enumerate}
    \item Regular (fast) rotators in the MAGNUS sample exhibit systematically higher specific angular momentum, quantified by \lmr, than non-regular (slow) rotators, consistent with trends observed in both the local Universe and at intermediate redshifts. Approximately 60\% of the fast rotators in our sample have a kinematic misalignment angle $\Psi_{\rm{mis}} < |10^{\circ}|$, a feature also common among local regular rotators. These results indicate that the internal dynamical structure of massive ETGs has remained stable since z $\sim$ 0.75.

    \item Based on the quantitative classification scheme of \citet{Cappellari_review_2016}, we found a slow rotator fraction  of 19.3$^{+3.0}_{-2.4}$\% (41 out of 213) in our sample, which is consistent with previous measurements at $z \sim 0.3$ and in the local Universe. The agreement across redshifts hints that the fraction of massive slow rotators has remained relatively constant over the last seven billion years.

    %\item We found that around 60\% of the fast rotators in the MAGNUS sample have misalignment angle $\Psi < |10^{\circ}|$, a trait also been observed for low redshift regular rotators suggesting kinematic structure of regular rotators have not changed dramatically over the last 5-6 billion years.

    \item The stellar population properties of the MAGNUS sample, for both SPS models, show a clear hierarchy in their dependence on $\sigma_{\mathrm{e}}$: metallicity $\langle[\mathrm{Z/H}]\rangle$ shows the strongest correlation, followed by $M_{\ast}/L$, while stellar age shows only a weak trend. We also observed that at fixed stellar mass, more compact galaxies tend to host older and more metal-rich stellar populations than their larger counterparts. These results are consistent with the archaeological downsizing picture, in which ETGs with higher velocity dispersions formed their stars earlier and reached a higher degree of chemical enrichment regulated by the depth of their gravitational potential wells. The observed scaling relations are consistent with findings from both local Universe and intermediate-redshift studies.
    
    % \item Similar to the slow rotator population in the local Universe, slow rotators in the MAGNUS sample tend to be more massive, exhibit higher $\sigma_\mathrm{e}$, and are more metal-rich with higher $M_{\ast}/L$ ratios compared to fast rotators.

    % \item A focused comparison between subsamples of slow and fast rotators, matched in velocity dispersion ($\sigma_\mathrm{e} >$ 200 \kmps), suggests that the higher metallicity and $M_{\ast}/L$ of slow rotators also correlate with their observed \lmr. This indicates that, beyond velocity dispersion, angular momentum plays a role in shaping stellar population properties.

    \item Slow rotators in the MAGNUS sample are preferentially more massive, have higher $\sigma_{\mathrm{e}}$, higher metallicity, and higher $M_{\ast}/L$ ratios than fast rotators, in agreement with the properties of the slow rotator population in the local Universe. A comparison between $\sigma_{\mathrm{e}}$-matched subsamples ($\sigma_{\mathrm{e}} > 200$ km s$^{-1}$) demonstrates that the metallicity difference is statistically significant, with median $\langle[\mathrm{Z/H}]\rangle$ of $0.12 \pm 0.02$ and $0.04 \pm 0.03$ for slow and fast rotators respectively, while a marginal but statistically insignificant age difference is also present. These results demonstrate that the kinematic state of ETGs, as quantified by \lmr, governs their stellar population properties beyond what is captured by velocity dispersion alone.
    
    %\item Comparing a sub-sample of slow rotators with a sub-sample of fast rotators within the MAGNUS sample, we found that the relatively higher metallicty and mass-to-light ratio of the slow rotators also depend on their kinematic state or \lmr when they have comparable velocity dispersions.

\end{enumerate}

In summary, the structural, kinematic, and stellar population scaling relations of the MAGNUS sample of massive ETGs at $0.25 < z < 0.75$ are consistent with both local Universe and intermediate-redshift studies, indicating that these relations were already in place by $z \sim 0.75$ and have remained stable over the last $\sim$7~Gyr. The galaxy properties and spatially resolved kinematics presented in this work will form the basis for subsequent papers in the MAGNUS series, including evolutionary analysis of rotational support, dynamical modeling, study of environmental influences, and the incorporation of kinematic constraints into time-delay cosmography.

\begin{acknowledgments}
P.M. and T.T. acknowledge support by NSF through grant AST-2407277.  C.D.F. acknowledges support for this work from the National Science Foundation under Grant No. AST-2407278.
\end{acknowledgments}


\vspace{5mm}
\textbf{Data availability}: The coordinates, redshifts, and measured structural, kinematic, and stellar population properties for a representative subset of the MAGNUS sample are presented in Table~\ref{tab:sample}. The full machine-readable catalog for all 213 galaxies is available in the online journal. \\

\facilities{VLT(MUSE), HST(ACS, WFC3)}


\software{
    \texttt{astropy} \citep{AstropyCollaboration13, AstropyCollaboration18,AstropyCollaboration2022},
    \texttt{jampy} \citep{Jampy_Cappellari_2008,Jampy_spherical_Cappellari_2020},
    \texttt{loess} \citep{ATLAS_XX_LOESS_Cappellari_2013b},
    \texttt{mgefit} \citep{MGE_Cappellari_2002},
    \texttt{mpdaf} \citep{MPDAF_Bacon_2016},
    \texttt{pafit} \citep{pafit_Krajnovic_2006},
    \texttt{ppxf} \citep{Cappellari_2004, Cappellari_2017_ppxf, LEGAC_ppxf_Cappellari_2023},
    \texttt{sep} \citep{SEP_Barbary_2016},
    \texttt{vorbin} \citep{Voronoi_Cappellari2003}
}



\appendix

\section{Sample Catalog}
\label{sec:sample_catalog}

\autoref{tab:sample} presents the measured and derived properties for a representative subset of the MAGNUS sample. Coordinates (RA, Dec) and redshifts are derived from the MUSE-DEEP datacubes (\autoref{sec:sample}). The SNR corresponds to the central spectrum extracted within one effective radius (\autoref{sec:stellar_kinematics}). Structural parameters, including  the ellipticity ($\epsilon$), semi-major axis ($R_{\rm e}^{\rm maj}$), and circularized effective radius ($R_{\rm e}$), are derived from MGE models of the HST imaging (\autoref{sec:photometry}). The integrated velocity dispersion ($\sigma_{\rm e}$) is measured from the central spectra (\autoref{sec:stellar_kinematics}). Specific angular momentum proxy (\lmr), is measured within an elliptical aperture of semi-major axis $2 R_{\rm e}^{\rm maj}$ as presented in \autoref{sec:lamr}. The photometric and kinematic position angles ($PA_{\rm phot}$ and $PA_{\rm kin}$) are described in \autoref{sec:photometry} and \autoref{sec:kin_PA}; the kinematic misalignment angle, $\Psi_{\rm mis}$, can be derived from these following Equation~(6) (\autoref{sec:kin_misalignment}). The kinematic classification (Class$_{\rm vis/\lambda_R}$) lists the visual and quantitative \lmr-based classification, respectively (\autoref{sec:result_kin_classifiation}). Stellar population properties, including luminosity-weighted age, metallicity, mass-to-light ratio, and stellar mass, are derived from the FSPS templates (\autoref{sec:sps_prop}). The full catalog of all 213 galaxies is available in machine-readable format in the online journal.

\begin{rotatetable*}
\begin{deluxetable*}{lcccccccccccccccc}
\setlength{\tabcolsep}{2.5pt}
\tablewidth{\textheight}
\tablecaption{Sample of measured and derived properties for the MAGNUS galaxies. The full table for all 213 galaxies is available in machine-readable format in the online journal.\label{tab:sample}}

\tablehead{
\colhead{ID} & \colhead{RA} & \colhead{Dec} & \colhead{$z$} & 
\colhead{SNR} & \colhead{$\epsilon$} & 
\colhead{$R_{\rm e}^{\rm maj}$} & \colhead{$R_{\rm e}$} & \colhead{$\sigma_{\rm e}$} &
\colhead{$\lambda_R$} & \colhead{$PA_{\rm phot}$} & 
\colhead{$PA_{\rm kin}$} & \colhead{Class$_{\rm vis/\lambda_R}$} & 
\colhead{$\langle \lg {\rm Age} \rangle$} & 
\colhead{$\langle [{\rm Z/H}] \rangle$} & 
\colhead{$M_{\ast}/L$} & \colhead{$\lg M_{\ast}$} \\
\colhead{} & \colhead{(deg)} & \colhead{(deg)} & \colhead{} & 
\colhead{(\AA$^{-1}$)} &  \colhead{} & 
\colhead{(kpc)} & \colhead{(kpc)} & \colhead{(km s$^{-1}$)} & \colhead{} & 
\colhead{(deg)} & \colhead{(deg)} & \colhead{} & 
\colhead{(yr)} & \colhead{(dex)} & 
\colhead{($M_{\odot}/L_{\odot}$)} & \colhead{($M_{\odot}$)}
}
\startdata
A1063\_39 & 342.184168 & -44.526333 & 0.351 & 26 & 0.33 & 1.46 & 1.182 & $122 \pm 5$ & $0.207 \pm 0.006$ & 51 & $48 \pm 10$ & FR/FR & 9.58 & -0.39 & 4.41 & 10.67 \\
A1063\_45 & 342.185532 & -44.518139 & 0.344 & 42 & 0.5 & 1.752 & 1.203 & $245 \pm 6$ & $0.207 \pm 0.004$ & 144 & $150 \pm 6$ & FR/FR & 9.78 & 0.16 & 5.03 & 11.13 \\
A1063\_51 & 342.18674 & -44.521944 & 0.339 & 16 & 0.5 & 0.994 & 0.703 & $140 \pm 7$ & $0.191 \pm 0.008$ & 163 & $178 \pm 30$ & U/FR & 9.99 & -0.17 & 5.43 & 10.59 \\
A1063\_58 & 342.18822 & -44.525389 & 0.35 & 36 & 0.52 & 1.09 & 0.755 & $276 \pm 7$ & $0.21 \pm 0.004$ & 150 & $148 \pm 8$ & FR/FR & 9.73 & 0.17 & 5.83 & 11.03 \\
A1063\_78 & 342.193441 & -44.517278 & 0.342 & 38 & 0.54 & 1.134 & 0.765 & $297 \pm 7$ & $0.2 \pm 0.003$ & 57 & $49 \pm 8$ & FR/FR & 9.82 & 0.16 & 5.22 & 11.04 \\
A1063\_85 & 342.195623 & -44.525444 & 0.346 & 48 & 0.25 & 2.279 & 1.976 & $229 \pm 4$ & $0.024 \pm 0.001$ & 4 & $36 \pm 30$ & SR/SR & 9.77 & 0.1 & 4.94 & 11.26 \\
A1063\_90 & 342.200376 & -44.524666 & 0.352 & 21 & 0.64 & 2.814 & 1.607 & $126 \pm 5$ & $0.341 \pm 0.008$ & 12 & $28 \pm 13$ & FR/FR & 9.91 & -0.28 & 4.5 & 10.73 \\
A1835\_1 & 210.267947 & 2.891601 & 0.262 & 42 & 0.19 & 1.658 & 1.489 & $201 \pm 4$ & $0.148 \pm 0.002$ & 60 & $60 \pm 9$ & FR/FR & 9.71 & 0.11 & 4.53 & 10.48 \\
A1835\_10 & 210.257434 & 2.864518 & 0.256 & 49 & 0.63 & 2.568 & 1.514 & $122 \pm 3$ & $0.49 \pm 0.004$ & 39 & $42 \pm 3$ & FR/FR & 9.54 & -0.32 & 3.09 & 10.83 \\
A1835\_11 & 210.254958 & 2.870045 & 0.26 & 35 & 0.32 & 1.961 & 1.605 & $136 \pm 4$ & $0.336 \pm 0.001$ & 34 & $38 \pm 6$ & FR/FR & 9.87 & -0.16 & 4.43 & 10.88 \\
A1835\_12 & 210.255487 & 2.86174 & 0.252 & 27 & 0.19 & 1.684 & 1.511 & $189 \pm 6$ & $0.401 \pm 0.004$ & 134 & $123 \pm 8$ & FR/FR & 9.89 & 0.07 & 5.08 & 11.05 \\
A1835\_2 & 210.2707 & 2.881379 & 0.244 & 53 & 0.15 & 2.949 & 2.735 & $186 \pm 4$ & $0.044 \pm 0.001$ & 96 & $84 \pm 30$ & SR/SR & 9.67 & -0.09 & 3.47 & 10.99 \\
A1835\_3 & 210.262607 & 2.879129 & 0.252 & 97 & 0.41 & 2.318 & 1.603 & $295 \pm 6$ & $0.344 \pm 0.001$ & 24 & $26 \pm 2$ & FR/FR & 9.8 & 0.16 & 4.76 & 11.42 \\
A1835\_4 & 210.25226 & 2.882295 & 0.25 & 28 & 0.65 & 0.573 & 0.339 & $207 \pm 12$ & $0.338 \pm 0.009$ & 21 & $8 \pm 20$ & FR/FR & 9.87 & 0.13 & 5.11 & 10.58 \\
A1835\_6 & 210.251092 & 2.879962 & 0.248 & 21 & 0.31 & 1.107 & 0.896 & $114 \pm 7$ & $0.159 \pm 0.004$ & 55 & $69 \pm 30$ & U/FR & 10.06 & -0.16 & 5.66 & 10.61 \\
\enddata
\tablecomments{Class$_{\rm vis/\lambda_R}$ lists the visual and quantitative $\lambda_R$-based kinematic classification, respectively (FR = fast rotator, SR = slow rotator, U = unclassified/inconclusive).}
\end{deluxetable*}
\end{rotatetable*}

\section{Kinematic maps of the MAGNUS Sample}
\label{sec:kinematic_maps}
\begin{figure*}
    \centering
    \includegraphics[width=\textwidth]{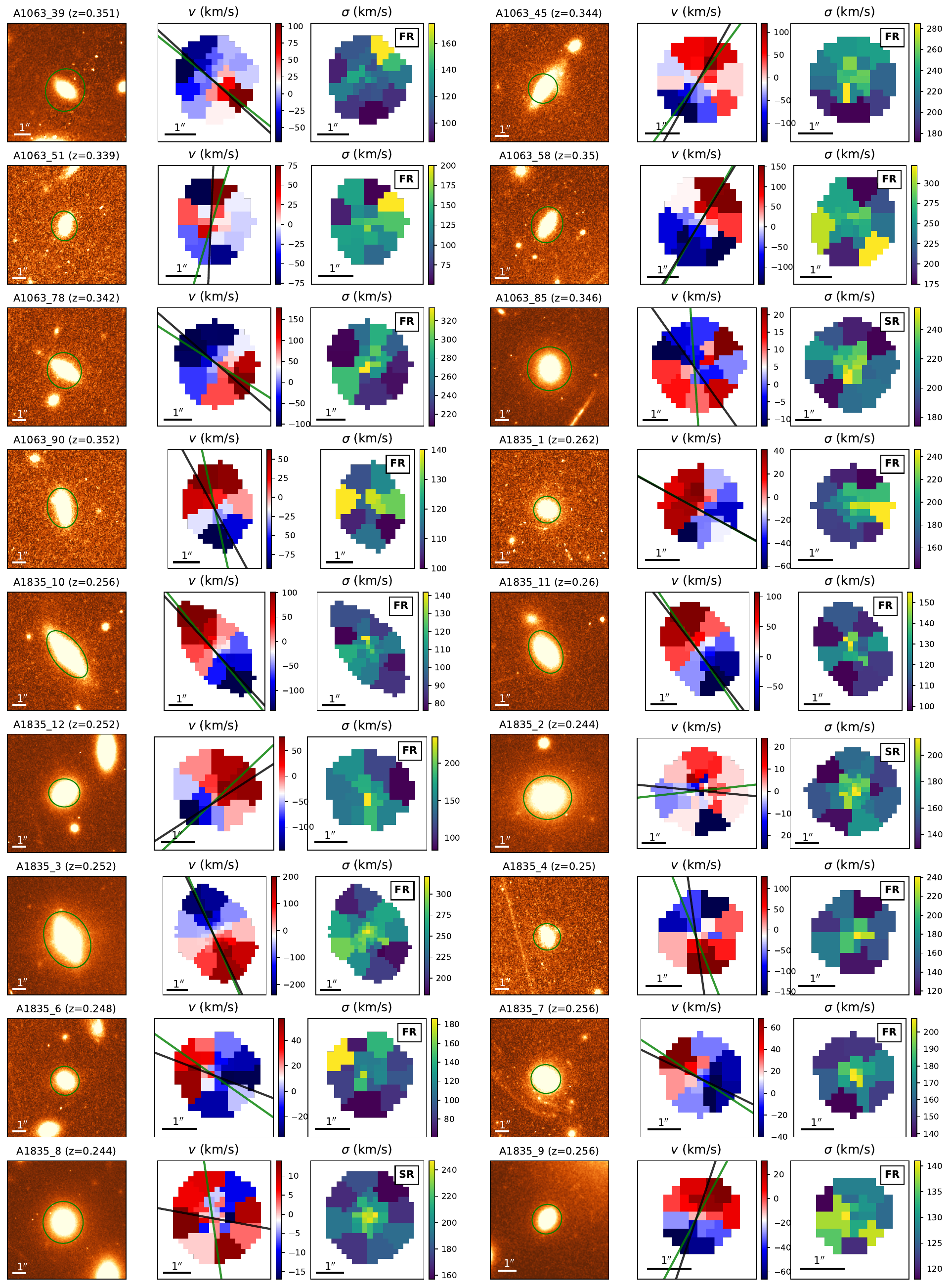}
    \caption{Representative figure from the full figure set showing kinematic maps for 18 galaxies. The full figure set (12 figures) is available in the online journal.}
    \label{fig:figset_1}
\end{figure*}

The HST image and the spatially resolved stellar kinematic maps (velocity, $v$, and velocity dispersion, $\sigma$) for the full MAGNUS sample are presented in this section. For each galaxy, three panels are shown, from left to right: the HST image, the median-subtracted velocity map, and the velocity dispersion map. The galaxy ID and redshift are given above each HST image, and the green curve outlines the region over which the kinematic 
data were extracted. The black and green lines in the velocity map indicate the kinematic and photometric major axes, respectively. The quantitative kinematic classification is also noted in the $\sigma$ panel (FR~=~fast rotator, SR~=~slow rotator). Color bars indicate the median-subtracted velocity and velocity dispersion in each Voronoi bin, and the white scale bar in each panel corresponds to $1''$. A representative subset of 18 galaxies is shown in \autoref{fig:figset_1}; the complete figure set for all 213 galaxies is available in the online journal.\\ \\

\figsetstart
\figsetnum{1}
\figsettitle{Kinematic Maps for 213 Galaxies}

\figsetgrpstart
\figsetgrpnum{1.1}
\figsetgrptitle{Galaxy Kinematics: Set 1}
\figsetplot{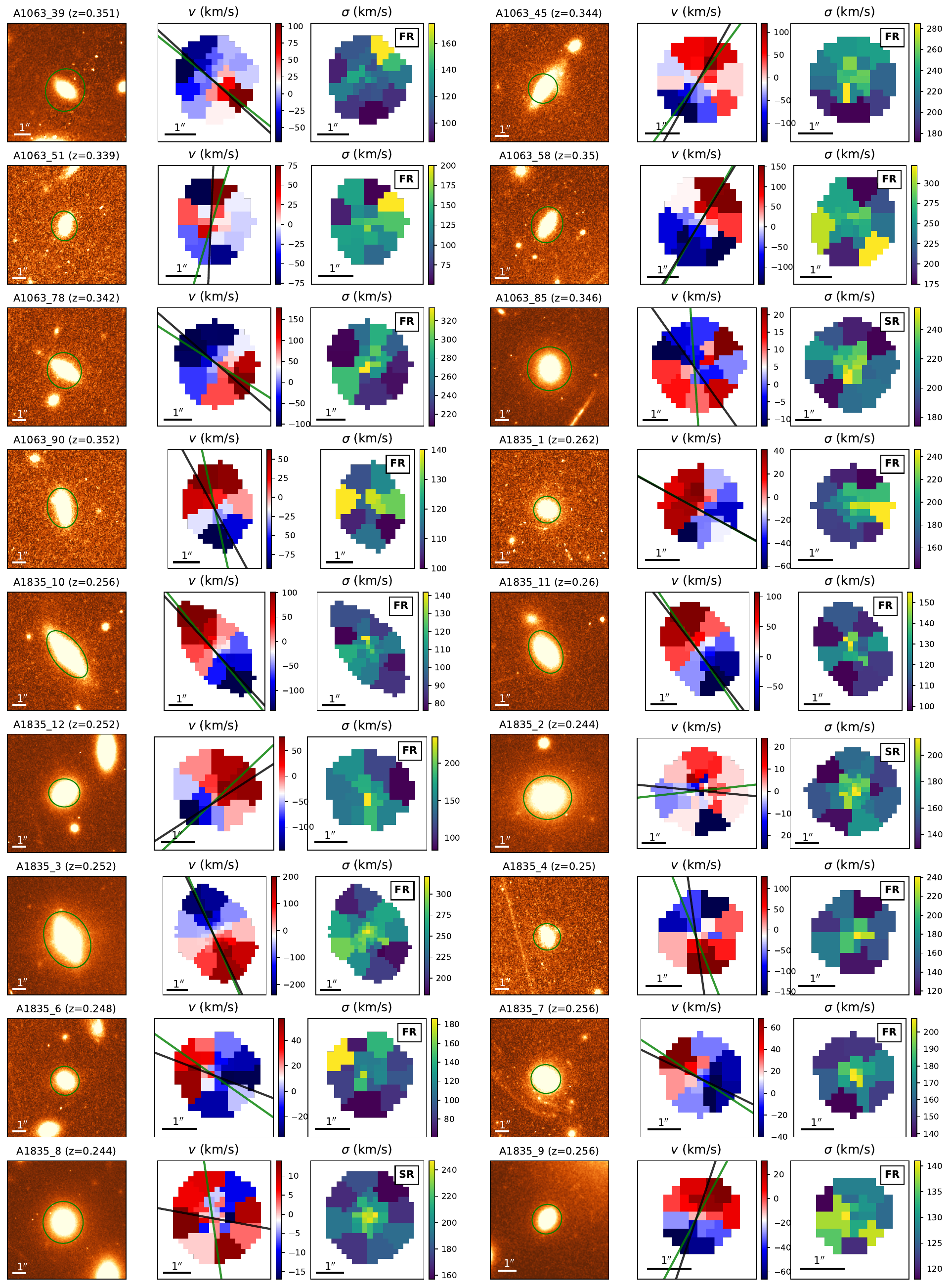}
\figsetgrpnote{HST image and stellar kinematic maps (see \autoref{sec:kinematic_maps} for panel description ) for the following 18 galaxies: A1063\_39, A1063\_45, A1063\_51, A1063\_58, A1063\_78, A1063\_85, A1063\_90, A1835\_1, A1835\_10, A1835\_11, A1835\_12, A1835\_2, A1835\_3, A1835\_4, A1835\_6, A1835\_7, A1835\_8, and A1835\_9.}
\figsetgrpend

\figsetend
% ----------- FIGURESET END --------------

% ----------- REPRESENTATIVE FIGURE -------------

\bibliographystyle{aasjournalv7}

\bibliography{Pritom_Mozumdar_2025}
 


\end{document}